\documentclass[12pt]{article} % Utilisez une classe de document standard
\usepackage[margin=2.5cm]{geometry} % Définir des marges de 2.5 cm de chaque côté
\usepackage{setspace} % Package pour gérer l'espacement entre les lignes
\usepackage{float}
\usepackage[utf8]{inputenc}

\usepackage{graphicx} %to import the images
\usepackage{multirow} % Required for multirow
\usepackage{rotating}
\usepackage{booktabs}
\usepackage{lscape}
\usepackage[titles,subfigure]{tocloft} % Alter the style of the Table of Contents
 \usepackage{amssymb}
 \newcommand{\bfs}[1]{\boldsymbol{#1}}
\usepackage{ulem}
\usepackage{color}
\usepackage{graphicx}
\usepackage{subcaption}
\usepackage[dvipsnames]{xcolor}

\newcommand{\KLB}[1]{\textcolor{black}{#1}}

\newcommand{\indep}{\perp \!\!\! \perp}

\usepackage[colorlinks,bookmarksopen,bookmarksnumbered,citecolor=red,urlcolor=red]{hyperref}

\usepackage{amsmath}
\usepackage{setspace}
\usepackage{enumitem}
\usepackage[numbers]{natbib} % Utiliser le package natbib avec l'option numbers pour des citations numérotées
\usepackage{graphicx}

\title{Continuous-time mediation analysis for repeatedly measured mediators and outcomes}
\author{
    Le Bourdonnec Kateline$^1$,
    Valeri Linda$^{2,3}$,
    Proust-Lima Cécile$^1$\\
    $^1$Univ. Bordeaux, Inserm, BPH, U1219, F-33000 Bordeaux, France\\
    $^2$Department of Biostatistics, Columbia University Mailman School of Public Health,\\
    722 W 168th St, New York, NY, USA\\
    $^3$Department of Epidemiology, Harvard T.H. Chan School of Public Health,\\
    677 Huntington Ave, Boston, MA, USA
}

\begin{document}

\begin{titlepage}
\author{Le Bourdonnec Kateline$^1$,Valeri Linda $^{2,3}$, Proust-Lima Cécile $^1$\\
$^1$Univ. Bordeaux, Inserm, BPH, U1219, F-33000 Bordeaux, France\\
$^2$ Department of Biostatistics, Columbia University Mailman School of Public Health,\\
722 W 168th St, New York, NY, USA\\
$^3$ Department of Epidemiology, Harvard T.H. Chan School of Public Health,\\
677 Huntington Ave, Boston, MA, USA}
\maketitle

\end{titlepage}
\section*{Abstract}
Mediation analysis aims to decipher the underlying causal mechanisms between an exposure, an
outcome, and intermediate variables called mediators. Initially developed for fixed-time mediator and outcome,
it has been extended to the framework of longitudinal data by discretizing the assessment times of mediator and outcome. Yet, processes in play in longitudinal studies are usually defined in continuous
time and measured at irregular and subject-specific visits. This is the case in dementia research when cerebral and cognitive changes measured at planned visits in cohorts are of interest. We thus propose a methodology to estimate the causal mechanisms between a time-fixed exposure ($X$),
a mediator process ($\mathcal{M}_t$) and an outcome process ($\mathcal{Y}_t$) both measured repeatedly over time in the presence
of a time-dependent confounding process ($\mathcal{L}_t$).
We consider \KLB{two} types of causal estimands, the natural effects and path-specific effects. We provide identifiability assumptions and \KLB{we employ a multivariate mixed model} based on differential equations for their estimation.
The performances of the method are assessed in simulations, and the method is illustrated in two real-world examples motivated by the 3C cerebral aging study to assess: (1) the effect of educational level on functional dependency through depressive symptomatology and cognitive functioning, and (2) the effect of a genetic factor on cognitive functioning potentially mediated by vascular brain lesions and confounded by neurodegeneration.\\
\textbf{Key words:} Causal Inference; Continuous-time; Dynamic model; Mediation Analysis;  Longitudinal data ;

\section{Introduction}

Mediation analysis, commonly used in public health, aims to decipher the underlying mechanism by which an independent variable ($X$) affects a dependent variable ($Y$) via one or more intermediate variables ($M$), also called mediators. The total effect between $X$ and $Y$ is split into a direct effect and indirect effects through the intermediate variables $M$.
Decomposing causal effects enhances our comprehension of the biological processes in play, and helps identify potential targets for therapeutic or prevention research.
%underlying causal mechanism between an exposure variable and an outcome variable, by discerning potential intermediate pathways and assessing the mediator's function.

The main framework for mediation analysis involves the definition of counterfactual outcomes in a hypothetical world. Counterfactual outcomes are unobserved variables corresponding to the value the outcome $Y$ would have taken if the exposure variable $X$ had been modified in a certain way. Causal effects can be defined as contrasts of counterfactual outcomes according to scenarios of intervention on $X$. Depending on the research question, different causal effects have been studied. For instance, the natural effects, introduced by \cite{robins_identifiability_1992-1}, contrast counterfactual outcome values had an individual been exposed at two distinct levels of exposure (i.e. values are set at the individual level). In contrast, the stochastic effects, introduced by  \cite{avin_identifiability_nodate}, contrast counterfactual outcome values had the distribution {of the mediator} been changed. 

Concepts and methods for mediation analysis have been primarily developed for exposure, mediator and outcome measured at a single time point. With the inherent dynamic nature of health processes, extensions to longitudinal data have been recently proposed. %Different approaches have been proposed in the literature to study survival outcomes. 
For instance, \cite{vanderweele_causal_2011} and \cite{lange_direct_2011} introduced methods for time-to-event outcomes with both the exposure variable and the mediator measured at a single time-point. Extensions to accommodate multiple mediators have followed \citep{lange_assessing_2014,huang_causal_2017}. More recently, \cite{vo_conduct_2020} proposed an alternative approach based on the natural effect proportional hazards model {for a single-time exposure on a survival-type outcome with} mediator and confounders both repeatedly measured over time. 
{Within the stochastic intervention approach, \cite{zheng_causal_2012} proposed a more flexible approach that handles exposure repeatedly measured over time and \cite{valeri_multistate_2023} considered time-to-events for both the mediator and outcome.}

In cerebral aging studies, the quantities of interest are usually dynamic processes such as cerebral volumes or cognitive functioning that are measured at planned visits. {Considering such repeated data structure is crucial to obtain accurate results in mediation analyses but it remains a challenge.}
\cite{vanderweele_mediation_2017} {defined} randomized interventional analogues to natural effects (i.e., a stochastic intervention on the mediator) for a repeated exposure variable, a repeated mediator and a fixed-time outcome. \cite{tai_causal_2023} extended the approach to multiple longitudinal mediators. Mediation analysis techniques to repeated outcomes and mediator data are nevertheless still limited. Some authors extended the mediation methods assuming processes evolve in discrete time with regular measures \citep{mittinty_longitudinal_2020,bind_causal_2016}. Yet, processes in play lie in continuous time and may be observed only sparsely at very irregular timings across individuals. \cite{albert_continuous-time_2019} considered a mediator and an outcome both defined in continuous time. They extended the identification assumptions of natural effects to continuous-time processes, and used a working model based on differential equations to estimate them. However, {this was achieved under the strong assumption} that {only the current mediator level} affects the outcome {when} the entire history of the mediator is likely involved. Moreover, the method did not consider time-varying confounders, and the estimation procedure was step-wise rather than {simultaneous for all the variables}  \citep{saunders_classical_2018}. 

 Motivated by applications in cerebral aging to decipher causal mechanisms among the multidimensional spheres involved, we propose in this work a continuous-time mediation analysis framework to estimate causal effects of time-fixed exposure ($X$) in a system of time-varying mediator ($\mathcal{M}_t$), counfounders ($\mathcal{L}_t$) and outcome ($\mathcal{Y}_t$), all defined in continuous-time and measured at irregular and sparse visits. We consider natural effects in the absence of time-varying confounders, and path-specific effects to accommodate the presence of time-varying confounding factors, {and we define for each case} the identifiability assumptions required for their estimation. Under weaker identifiability assumptions, randomized interventional analogues to natural effects can also be estimated under the proposed strategy. Our procedure relies on a recently developed 
%Few methodologies are currently able to deal with the complexity of ageing, in particular the multiplicity of dynamic processes to be investigated, such as cerebral vascular lesions, neurodegeneration and cognitive and functional declines. Understanding the cause-and-effect relationships between these dimensions, and the causal mechanisms of certain potential risk factors on this succession of impairments, remains an open question. However, current methods do not consider the time-dependent aspect of these impairments. To overcome this limitation, Taddé et al. (2019) proposed recently 
multivariate mixed model \citep{tadde_dynamic_2020} that quantifies the influences between dynamic processes of a network to estimate the conditional distributions of the mediator, confounder, and outcome processes from which the mediation g-formulas can be applied and the causal contrasts estimated. Two applications in cerebral aging research are considered from the population-based 3C cohort \citep{3c_study_group_vascular_2003}. We first estimate the pathways through cognitive functioning and depressive symptomatology involved in the association between educational level and functional dependency among olders. We secondly assess the relationship between $\epsilon_4$ allele of the apolipoprotein $E$ (APOE4), the main genetic factor for dementia, and cognitive functioning exploring the pathways through vascular cerebral lesions and neuro-degeneration. 

\section{Methods}
\subsection{Notation}
\begin{figure}[h]
  \centering  
  \begin{minipage}[t]{0.45\textwidth}
    \centering
    \caption*{(A) Without time-varying confounders}
    \includegraphics[width=1\linewidth, height=6cm]{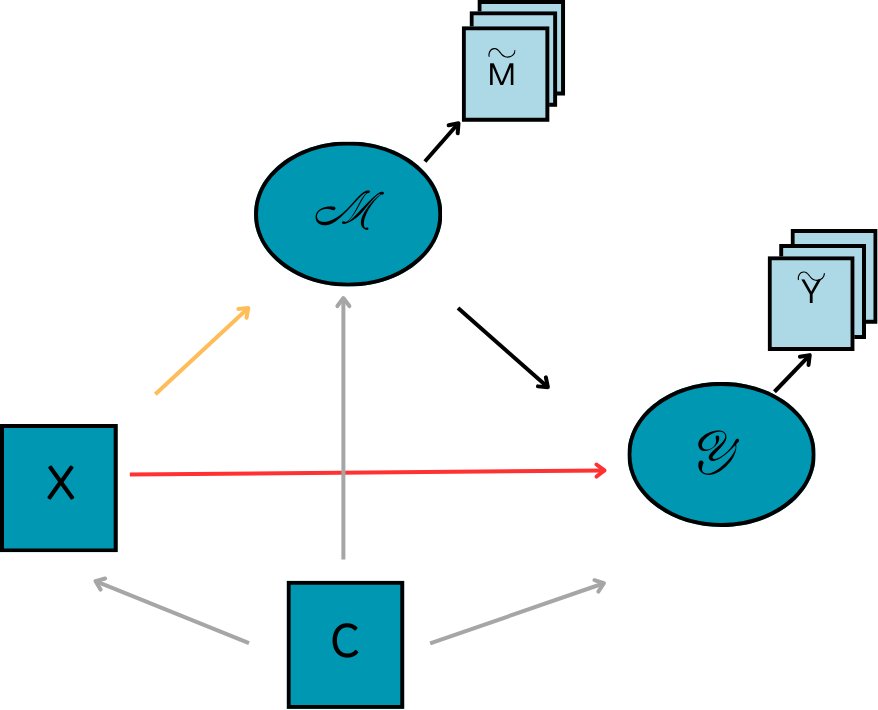} 
  \end{minipage}\hfill
  \begin{minipage}[t]{0.6\textwidth}
    \centering
     \caption*{(B) With time-varying confounders}
    \includegraphics[width=1\linewidth, height=8.7cm]{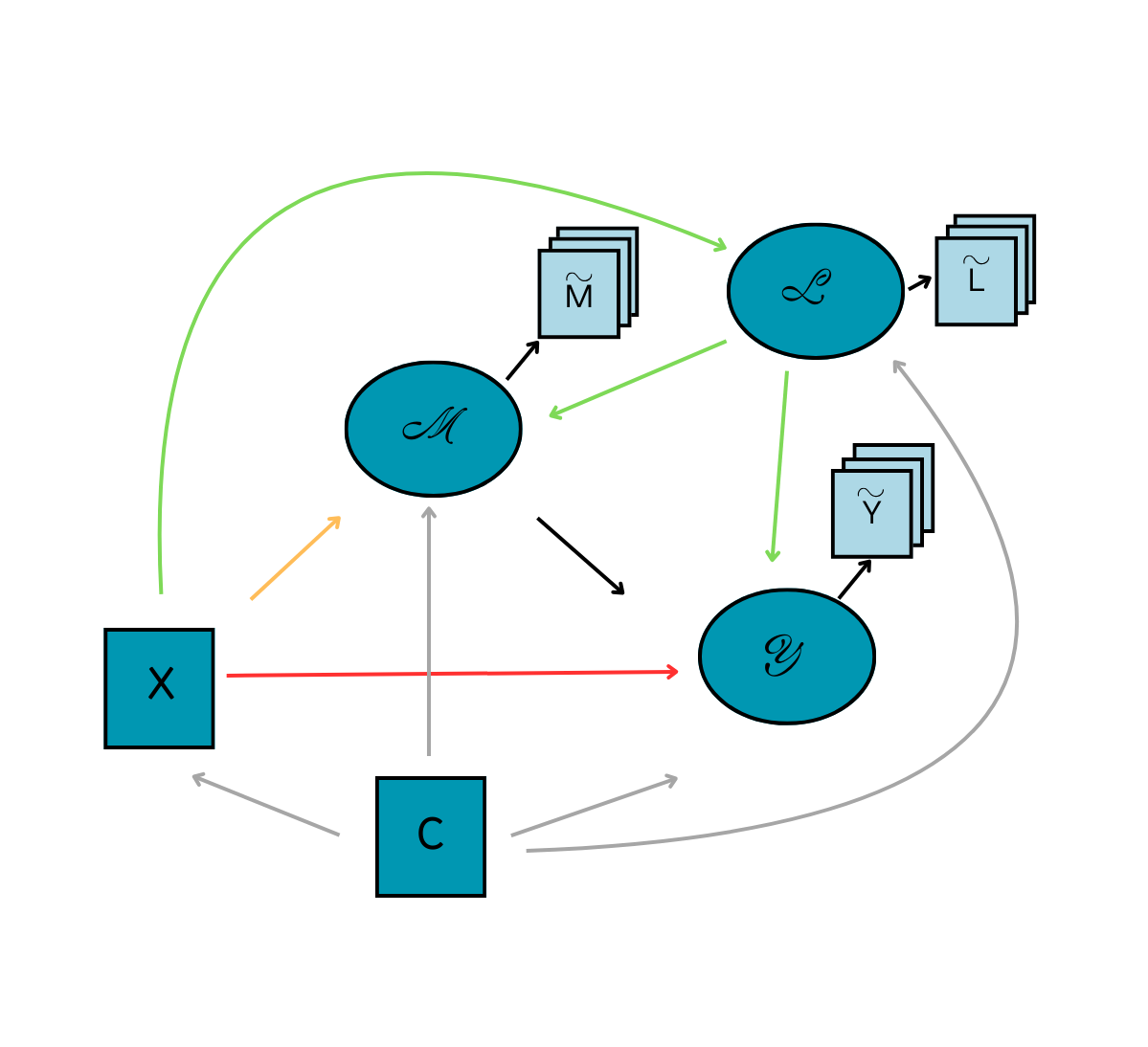} 
  \end{minipage}
  \caption{Causal Mediation Path Diagram exploring the mechanism between a time-fixed exposure X and an outcome process $\mathcal{Y}_t$ through (A) mediator process $\mathcal{M}_t$ or (B) mediator $\mathcal{M}_t$ and time-varying confounders $\mathcal{L}_t$ processes, given baseline confounders C. A local dependence graph approach is used for relating the processes altogether. In particular, the absence of an arrow from process $\mathcal{Y}$ to process $\mathcal{M}$ postulates no effect of the history of the outcome on future values of the mediator. Ovals indicate latent processes while squares indicate observed variables}
 \label{tab:Fig1}
\end{figure}

We consider the setting described in Figure \ref{tab:Fig1} with a time-fixed exposure $X$ (defined in $\{0, 1\}$ or $\mathbb{R}$), time-fixed confounders $\bfs{C}$ (with $C \in \mathbb{R}^P$), as well as a time-dependent mediator and a time-dependent outcome, possibly in the presence of time-dependent confounders $L$ (right panel). The time-dependent mediator, confounder, and outcome are processes defined in continuous time, with their values at any time $t$ ($t \in \mathbb{R}^+$) denoted as $M(t)$, $Y(t)$, and $L(t)$, respectively, and belonging to $\mathbb{R}$. Their history up to time $t$ is denoted by $\mathcal{M}_t$, $\mathcal{Y}_t$, and $\mathcal{L}_t$. In a cohort study, this setting translates into the collection of the exposure $X_i$ and the confounder $C_i$ at baseline for each subject $i$ ($i = 1, \dots, N$), and prone-to-error measures of the mediator $\tilde{M}_{ij}$, the outcome $\tilde{Y}_{ij}$, and the time-dependent confounders $\tilde{L}_{ij}$ at discrete visits $j$ ($j = 1, \dots, n_i$). These visits usually occur at subject-specific times $t_{ij}$, potentially with different time schedules across the time-dependent variables. In our setting, we systematically postulate that the outcome does not affect future values of the intermediate processes, and that the mediator does not affect future values of the confounder process.

To define causal effects, we introduce counterfactual variables. The counterfactual outcome $Y_t(x,\mathcal{L}_t(x''), \mathcal{M}_t(x',l_t))$ is the value \KLB{the outcome would have taken at time $t$ had $X$ been} set to $x$, had $\mathcal{L}_t$ been set to the value it would have taken \KLB{had $X$ been} set to $x''$, and had $\mathcal{M}_t$ been set to the value it would have taken had $X$ been set to $x'$ and had $\mathcal{L}_t$ been set to $l_t$. 

\subsection{Causal effects definition}

Depending on the framework and the objective, different causal effects can be defined using potential outcomes. We primarily consider natural effects, both in the absence of time-varying confounders, and in the presence of time-varying confounders with path-specific effects. \KLB{Stochastic intervention analogues are mentioned in Discussion}.

\subsubsection{Natural effect} \label{defNatural}

The natural effect decomposes the total effect (TE) of the exposure $X$ on the outcome at time $t$, $Y(t)$, into a natural direct effect (NDE) defined as the effect of $X$ on $Y(t)$ only, and a natural indirect effect (NIE) defined as the effect of $X$ on $Y(t)$ passing through the mediator process up to $t$, $\mathcal{M}_t$. 

Contrasting two levels $x$ and $x'$, these effects are expressed as follows:
\begin{align}
 &TE = \mathbb{E}(Y_t(x,\mathcal{M}_t(x))|C) - \mathbb{E}(Y_t(x',\mathcal{M}_t(x'))|C)
 \label{NE}\\
 &NIE = \mathbb{E}(Y_t(x,\mathcal{M}_t(x))|C) - \mathbb{E}(Y_t(x,\mathcal{M}_t(x'))|C)\label{NIE}\\
 &NDE = \mathbb{E}(Y_t(x,\mathcal{M}_t(x'))|C) - \mathbb{E}(Y_t(x',\mathcal{M}_t(x'))|C)
\label{NDE}
\end{align}

\noindent where $TE$ is the difference in outcome \KLB{at time $t$} had an individual been exposed at level $x'$ instead of $x$, $NIE$ is the difference in outcome \KLB{at time $t$} had an individual been exposed to level $x$ while the mediator process \KLB{up to $t$} changed from what observed under level $x'$ compared to level $x$. The $NDE$ is the difference in outcome \KLB{at time $t$} had an individual been exposed two different levels $x$ and $x'$, while keeping the mediator process \KLB{up to $t$} fixed at what would be under exposure level $x'$. \KLB{These effects are marginal effects. They incorporate the pathways through the history of the outcome up to $t$. The history of the outcome that would be affected by the history of the mediator is part of the indirect effect, the one not affected by the history of the mediator is part of the direct effect.}

%Replacing $\mathcal{M}_t$ with stochastic intervention $G_{\mathcal{M}_t}$ {in equations \eqref{NE}, \eqref{NIE}, \eqref{NDE}}, randomized interventional analogues to natural effects can be identified under  weaker assumptions \citep{vanderweele_mediation_2017}. 

\subsubsection{Path-specific effect} \label{defPath}

In the presence of time-dependent confounders, the path-specific effect approach decomposes the effect of the exposure $X$ on the outcome at time $t$, $Y(t)$, into the direct effect of $X$ on $Y(t)$ (noted  PSE$_{XY}$), the effect passing only through the mediator process up to $t$, $\mathcal{M}_t$ (noted PSE$_{XMY}$), and the effect passing first through the time-dependent confounder process up to $t$, $\mathcal{L}_t$ (noted PSE$_{XL(M)Y}$). \KLB{Again, we consider marginal effects.}

Contrasting two levels $x$ and $x'$, these effects are expressed as follows:
\begin{align}
&PSE_{XY} = \mathbb{E}(Y_t(x,\mathcal{L}_t(x'), \mathcal{M}_t(x',\mathcal{L}_t(x'))|C) -\mathbb{E}(Y_t(x',\mathcal{L}_t(x'), \mathcal{M}_t(x',\mathcal{L}_t(x'))|C)
\label{PSE1}\\
&PSE_{XMY} = \mathbb{E}( Y_t(x,\mathcal{L}_t(x'), \mathcal{M}_t(x,\mathcal{L}_t(x'))|C) - \mathbb{E}(Y_t(x,\mathcal{L}_t(x'), \mathcal{M}_t(x',\mathcal{L}_t(x'))|C)
\label{PSE2}\\
& PSE_{XL(M)Y} = \mathbb{E}( Y_t(x,\mathcal{L}_t(x), \mathcal{M}_t(x,\mathcal{L}_t(x))|C) -\mathbb{E}( Y_t(x,\mathcal{L}_t(x'), \mathcal{M}_t(x,\mathcal{L}_t(x'))|C)
\label{PSE3}
\end{align}

For example, $PSE_{XMY}$ (equation \eqref{PSE2}) is the expected difference in outcome \KLB{at time $t$} had the mediator process only changed due to an exposure change from $x'$ to $x$, while keeping fixed the exposure and the time-dependent confounder process.

\subsection{Assumptions} \label{assump}

The causal contrasts and expectations defined above rely on potential outcomes that are not observable.
%To identify these effects from the observations, it is mandatory to comply to foursets of fundamental assumptions:consistency, positivity, sequential ignorability and cross-world independence.
To identify these effects from observations, four basic assumptions must be made: \textbf{consistency, positivity, sequential ignorability} and \textbf{cross-world independence}.
These assumptions vary depending on the presence or absence of time-varying confounders. 
In the following, we make the distinction between the two cases with $\overline{TVC}$ and $TVC$ specifying the absence and the presence of time-varying confounding factors.

%In the following, we distinguish between the two cases with $\overline{TVC}$ and $TVC$ indicating the absence and the presence of time-varying confounders, respectively. 

\begin{enumerate}[label=(\roman*)]
    \item  $\textbf{consistency}$:
This assumption establishes the connection between counterfactual variables and observed variables \citep{nguyen_clarifying_2022}. The value of the observed outcome ${Y}(t)$ is equal to the value of the corresponding counterfactual outcome, i.e.     %\begin{enumerate}[label=\textbullet]
        %\item $\overline{TVC}$: ${Y}(t)$=$Y_t(x, m_t) $ if X=x, $\mathcal{M}_t=m_t$
        %\item TVC:  $Y(t)=Y_t(x, m_t,l_t) $ if X=x, %$\mathcal{M}_t=m_t$ and $\mathcal{L}_t=l_t$
 %   \end{enumerate} 
 for $\overline{TVC}$: ${Y}(t)$=$Y_t(x, m_t) $ if X=x, $\mathcal{M}_t=m_t$; and for TVC:  $Y(t)=Y_t(x, m_t,l_t) $ if X=x, $\mathcal{M}_t=m_t$ and $\mathcal{L}_t=l_t$.

    \item \textbf{positivity} : 
The positivity assumption stipulates that each individual, given their values of $C$, has a positive probability of receiving any exposure value, and a positive probability to have any mediator and time-varying confounder history \citep{nguyen_clarifying_2022}.
%Given $C$, there is a non-zero probability that the exposure $X$, takes any value $x$ 
It means that for $\overline{TVC}$: $P(X=x|C) > 0$ %,   $P(\mathcal{L}_t(x)=l_t|X=x,C) > 0 $ 
and {$P(\mathcal{M}_t(x)=m_t|X=x,C) > 0 $}, and for TVC: $P(X=x|C) > 0 $, $P(\mathcal{L}_t(x)=l_t|X=x,C) > 0 $ and $P(\mathcal{M}_t(x,l_t)=m_t|X=x,C,\mathcal{L}_t=l_t) > 0 $. 
    \item \textbf{sequential ignorability}: this assumption defines the absence of unobserved confounding in the system through 3 to 5 sub-assumptions:
    \begin{enumerate}
        \item There is no unobserved confounding of the effect of $X$ on $Y_t$ given other variables, i.e.: $Y_t(x, m_t) \indep X|C$, for $\overline{TVC}$, and $Y_t(x, l_t, m_t) \indep X|C$ for TVC.
        
        \item There is no unobserved confounding of the effect of intermediate processes on $Y_t$  given $C$ and $X$, i.e., $Y_t(x, m_t) \indep \mathcal{M}_t|C,X$ for $\overline{TVC}$, and $Y_t(x, l_t, m_t) \indep  \mathcal{L}_t,  \mathcal{M}_t|C,X$ 
        for TVC.
        
        \item There is no unobserved confounding of the effect of $X$ on intermediate processes given C, i.e., $\mathcal{M}_t(x) \indep X|C$ for $\overline{TVC}$, and $\mathcal{L}_t$,$\mathcal{M}_t(x) \indep X|C$ for TVC.
        
        \item For TVC only, there is no unobserved confounding between the mediator process $\mathcal{M}_t$ and (X,$\mathcal{L}_t$) given C, i.e., $\mathcal{M}_t(x,l_t) \indep X,\mathcal{L}_t |C$.
        \item For TVC only, there is no unobserved confounding between the counfounder process $\mathcal{L}_t$ and the mediator process $\mathcal{M}_t$ affected by X given C, i.e., $\mathcal{M}_t(x,l_t) \indep \mathcal{L}_t(x')|X,C$.
       \end{enumerate}
   \item \textbf{Cross-world independence assumption:} this assumption stipulates the absence of unobserved confounding in the counterfactual world:
   \begin{enumerate}
       \item 
       %There is no unobserved confounding between the potential outcome and intermediates variables, 
       \KLB{No confounding between the outcome and the joint intermediate variables processes that is affected by the exposure} i.e., $Y_t(x,m_t) \indep \mathcal{M}_t(x')|X,C$ for $\overline{TVC}$, and $Y_t(x,l_t,m_t) \indep \mathcal{L}_t(x'),\mathcal{M}_t(x')|X,C$ for TVC.

       \item For TVC only, there is no confounding between outcome and each of intermediate processes \KLB{that is affected by the exposure}, i.e., $Y_t(x,l_t,m_t) \indep \mathcal{L}_t(x'),\mathcal{M}_t(x\KLB{''},l_t)|X, C$.
\end{enumerate}
\end{enumerate}

\subsection{Identification}

\subsubsection{Natural Effects}\label{NEs}

The natural effects are defined as a comparison of two expectations $\upsilon = \mathbb{E}({Y}_t(x,\mathcal{M}_t(x'))|C)$ where $x$ and $x'$ can take various values depending on the effect (NE, NIE, NDE). Using the assumptions ($(i), (ii), (iii.a), (iii.b), (iii.c), (iv.a)$), this expectation can be written as a function of the observations, and thus becomes identifiable. 
Thanks to assumption \textit{(iii.a)}, we can rewrite  $\upsilon =\mathbb{E}({Y}_t(x,\mathcal{M}_t(x'))|C)$ in $\upsilon =\mathbb{E}({Y}_t(x,\mathcal{M}_t(x'))|C,X)$.
First, the estimand can be developed according to the mediator history $\mathcal{M}_t$ :
\begin{align*}
    \upsilon =& \int_{m_t}^{} \mathbb{E}({Y_t}(x,m_t)|C=c,X=x, \mathcal{M}_t(x')=m_t) \times f_{\mathcal{M}_t(x')| C=c}(m_t)~ d_{\mathcal{M}_t(m_t)}
\end{align*}

Second, thanks to assumption ($iv.a$), we can remove the conditioning $\mathcal{M}_t(x')=m_t$ and in order to use the consistency assumption, we add $X=x'$ in the density of $\mathcal{M}_t$ thanks to assumptions ($i$) and ($iii.c$):
\begin{align*}
   \upsilon =& \int_{m_t}^{} \mathbb{E}({Y_t}(x,m_t)|C=c, X=x) \times f_{\mathcal{M}_t(x')|C=c, X=x'}(m_t)~ d_{\mathcal{M}_t(m_t)}
\end{align*}

%\begin{align*}
%    \upsilon =& \int_{m_t}^{} \mathbb{E}({Y_t}(x,m_t)| C=c, X=x)  \times f_{\mathcal{M}_t(x')|C=c}(m_t)~ d_{\mathcal{M}_t(m_t)}
%\end{align*}

%Third, in order to use the consistency assumption, we add $X=x'$ in the density of $\mathcal{M}_t$ thanks to assumptions ($i$) and ($iii.c$):

We add the conditioning on $\mathcal{M}_t=m_t$ in the expectation of $Y_t$ thanks to assumption ($iii.b$): 
\begin{align*}
   \upsilon = \int_{m_t}^{} \mathbb{E}({Y_t}(x,m_t)|C=c,X=x,\mathcal{M}_t=m_t) \times f_{\mathcal{M}_t|C=c, X=x'}(m_t)~ d_{\mathcal{M}_t(m_t)}
\end{align*}

Finally, the consistency assumption $i$ can be applied to obtain an expression that only involves the observations so that:
\begin{align}
\mathbb{E}(Y_t(x,\mathcal{M}_t(x'))|C) = &\int_{m_t}^{}\mathbb{E}(Y_t|C=c,X=x,\mathcal{M}_t=m_t)\times f_{\mathcal{M}_t|C=c, X=x'}(m_t)~ d_{\mathcal{M}_t(m_t)}
\label{expectNE}
\end{align}

\subsubsection{Path-specific effect}\label{PSE}

In the presence of time-varying confounders $\mathcal{L}_t$, assumption $iii.c$ is violated and the third step of the identification of the natural effect cannot be applied. Alternatively, path-specific effects are considered. They are defined as a comparison of two expectations $\xi = \mathbb{E}({Y}_t(x,\mathcal{L}_t(x),\mathcal{M}_t(x',\KLB{\mathcal{L}_t(x)}))|C)$, where $x$ and $x'$ can take various values. 

Using similar developments as for the natural effects (see Web Supplementary Material, SECTION 1 for details), we obtain: 
\begin{equation}
\begin{split} \label{expectPSE}
\xi =  \int_{l_t}^{} \int_{m_t}^{} & \mathbb{E}(Y_t|C=c, X=x, \mathcal{L}_t=l_t, \mathcal{M}_t=m_t) \times  \\
& f_{\mathcal{L}_t|C=c, X=x}(l_t) \times f_{\mathcal{M}_t|(C=c, X=x', \mathcal{L}_t=l_t)}(m_t) ~ d_{m_t} d_{l_t} 
    \end{split}
\end{equation}

%where a different set of assumptions is used depending on the path: 
\KLB{with different assumptions for each path:}
assumptions ($i$)-($iii.c$) and ($iv.a$) for the $PSE_{XY}$ (direct path), assumptions ($i$)-($iii.c$),($iii.e$),($iv.b$) for the $PSE_{XMY}$ (path through $\mathcal{M}_t$), and assumptions ($i$)-($iii.a$), ($iii.c$), ($iv.a$)-($iv.b$) for the $PSE_{X(L)MY}$ (indirect paths through $\mathcal{L}_t$).  

\subsection{Estimation}

\subsubsection{Monte Carlo approximation \KLB{(MCA)}}\label{monte_carlo}
  
The causal contrasts are differences of expectations from Equations \eqref{expectNE} and \eqref{expectPSE} for the natural and the path-specific effects, respectively. These expectations are general expressions to be estimated from the data. In specific cases, an analytical solution can be computed. Otherwise, the integrals can be approximated by the Monte Carlo approach with $B$ Monte-Carlo replicates. For the PSE, this gives:
\begin{align*}
 &\mathbb{E}(Y_t(x,\mathcal{L}_t(x'), \mathcal{M}_t(x',\mathcal{L}_t(x'))|\KLB{C})
        \approx \sum_{k=1}^{B}~\mathbb{E}(Y_t|X=x,\mathcal{L}_t=l_{t}^{(k)},\mathcal{M}_t=m_{t}^{(k)}, \KLB{C})
\end{align*}
 with random draws $l_{t}^{(k)} \sim f_{\mathcal{L}_t| X=x', C=c}$ and $m_{t}^{(k)} \sim f_{\mathcal{M}_t|C=c, X=x', \mathcal{L}_t=l_{tj}}$ from the conditional distributions. The same approximation is obtained for the natural effects by removing mention to the TVC.
 %The Monte-Carlo approximation highlights that the causal estimands calculation requires:
 \KLB{The MCA highlights that the calculation of causal estimators relies on certain quantities, estimated using a statistical model, called the working model.}
\begin{itemize}
    \item the conditional expectation of $Y_t$ given the intermediate processes and time-fixed factors;
    \item the conditional distribution of $\mathcal{M}_t$ given the exposure, (and the time-varying confounders);
    \item in the presence of $\mathcal{L}_t$, their conditional distribution on $X$ and the time-fixed confounders.
\end{itemize} 
%\KLB{These quantities are to be estimated from a statistical model, called working model.}

\subsubsection{Example of working model}\label{working_m}

\KLB{Any statistical model can be used to estimate these causal effects as long as (i) the processes are jointly modelled, taking into account the association structures both between processes and with the time-fixed exposure, (ii) the model handles the data characteristics such as irregular timings or missing data mechanisms, and (iii) the conditional distributions of the processes can be derived.} In this work, we opted for a multivariate mixed model based on differential equations \citep{tadde_dynamic_2020}. 
%This model was developed to quantify the temporal influences between a system of latent processes measured by repeated marker data 
\KLB{This model quantifies the temporal influences between latent processes measured by repeated marker data under a Missing At Random (MAR) mechanism}. In our case, the processes are  $\mathcal{L}_t, \mathcal{M}_t$ and $\mathcal{Y}_t$ with values $L_i(t)$, $M_i(t)$ and $Y_i(t)$ at time $t$ {for subject $i$}  ($i=1,...,N$). Their trajectories are defined by the initial level at time 0 and the instantaneous change over time, both modelled in the mixed modeling framework using random effects \citep{laird_random-effects_1982}. The temporal dependencies between the processes are modelled by the effect of the current level of one process on the instantaneous change of another. \KLB{By modeling the instantaneous change of processes over time, the working model thus allows the mediator and outcomes trajectories to be functions of the history of the intermediate processes.} The working model is formulated as follows:  
\begin{align}
\begin{aligned}
    & \text{For process } \mathcal{L}_t: \left\{
    \begin{array}{ll}
    \begin{aligned}
         L_i(0)  &= \textbf{X}_i^{L(0)} \bfs{\beta}^L + {u}_i^L \\
    \frac{\partial L_i(t)}{\partial t} &= \textbf{X}_i^{L}(t)\bfs{\gamma}^L + \textbf{Z}_i^L(t)\textbf{v}_i^L
    \end{aligned}
    \end{array}\right. \\
    & \text{For process } \mathcal{M}_t: \left\{
    \begin{array}{ll}
    \begin{aligned}
    M_i(0) &= \textbf{X}_i^{M(0)}\bfs{\beta}^M + {u}_i^M \\
    \frac{\partial M_i(t)}{\partial t} &= \textbf{X}_i^{M}(t)\bfs{\gamma}^M + \textbf{Z}_i^M(t)\textbf{v}_i^M + \alpha^{ML}_i L_i(t)
    \end{aligned}
    \end{array}\right. \\
    & \text{For process } \mathcal{Y}_t: \left\{
    \begin{array}{ll}
    \begin{aligned}
    Y_i(0) &= \textbf{X}_i^{Y(0)}\bfs{\beta}^Y + {u}_i^Y \\
    \frac{\partial Y_i(t)}{\partial t} &= \textbf{X}_i^{Y}(t)\bfs{\gamma}^Y + \textbf{Z}_i^Y(t)\textbf{v}_i^Y + \alpha^{YL}_i L_i(t) + \alpha^{YM}_i M_i(t)
    \end{aligned}
    \end{array}\right.
\end{aligned}
 \label{structworkmod}
\end{align}

where $\textbf{X}_i^{L(0)}$, $\textbf{X}_i^{M(0)}$, $\textbf{X}_i^{Y(0)}$ are the vectors of covariates (including intercept, $X$ and $\bfs{C}$) associated with the initial levels of the three processes through parameters 
$\bfs{\beta}^L$, $\bfs{\beta}^M$, $\bfs{\beta}^Y$. Vectors of covariates $\textbf{X}_i^{L}(t)$, $\textbf{X}_i^{M}(t)$, $\textbf{X}_i^{Y}(t)$ are associated with the change over time of the processes through parameters $\bfs{\gamma}^L$, $\bfs{\gamma}^M$, $\bfs{\gamma}^Y$. They include the intercept, time functions (allowing for nonlinear change over time), the exposure and confounders as well as their eventual interactions with the time functions. The vectors $\textbf{Z}_i^L(t)$, $\textbf{Z}_i^M(t)$, $\textbf{Z}_i^Y(t)$ include the intercept and, eventually time functions, to be associated with the changes over time through the individual random effects $\textbf{v}_i^L$, $\textbf{v}_i^M$, $\textbf{v}_i^Y$. The random intercepts on the initial levels ${u}_i^L$, ${u}_i^M$, ${u}_i^Y$, and the random effects on the changes over time $\textbf{v}_i^L$, $\textbf{v}_i^M$, $\textbf{v}_i^Y$ \KLB{define the vector of individual random effects that account for the intra-individual correlation. It follows a zero-mean multivariate normal distribution, with a semi-structured variance matrix: within each process, ${u}_i^{~.}$ and $\textbf{v}_i^{~.}$ are correlated but across processes, only the ${u}_i^{~.}$ are correlated the one with the other.} The influences between processes are captured by the $\alpha^{aa'}_i$ that quantify the effect of process $a'$ on the instantaneous change over time of process $a$. The influences $\alpha^{aa'}_i$ can be modelled as a linear combination of covariates to account for instance for exposure interaction with intermediate processes: $\alpha^{aa'}_i = \alpha_0^{aa'} + {\bfs{X}_i^\top} \bfs{\alpha}_1^{aa'}$  with $\bfs{X_i}$ a vector of covariates. 

The structural models at the process level are linked to the error-prone observations in equations of observations, assuming in this work additive errors. 
: 
\begin{equation}
    \left\{
    \begin{array}{lll}
         \tilde{L}_{ij} = L_i(t_{ij}) + \epsilon_{ij}^L & & \text{ for } j=1,...,n_{i}^L\\
         \tilde{M}_{ij} = M_i(t_{ij}) + \epsilon_{ij}^M & & \text{ for } j=1,...,n_{i}^M\\
         \tilde{Y}_{ij} = Y_i(t_{ij}) + \epsilon_{ij}^Y& &  \text{ for } j=1,...,n_{i}^Y\\
    \end{array}\right.\label{obsworkmod}
\end{equation}

with independent zero-mean Gaussian errors $\epsilon_{ij}^L$, $\epsilon_{ij}^M$, and $\epsilon_{ij}^Y$ of variances $\sigma_L^2$, $\sigma_M^2$, and $\sigma_Y^2$.

 \KLB{Let denote $\theta$ the total vector of parameters of the working model, containing the regression parameters in Equation \eqref{structworkmod} and the variance parameters of the random effects and errors.} The model is estimated in the maximum likelihood framework in the R package CInLPN (https://github.com/bachirtadde/CInLPN). To achieve an analytical likelihood calculation, the program approximates the differential equations by difference equations with a fine time grid defined by a step $\delta$ to be specified by the user. Since this model is Gaussian and linear, analytical solutions can be found for \KLB{ the expectations $\upsilon$ and $\xi$ and the counterfactual contrasts defined in subsections \ref{NEs} and \ref{PSE}} without requiring a \KLB{MCA}.

\subsubsection{Confidence intervals}

Since the model is parametric, the confidence intervals of \KLB{the counterfactual contrasts defined in subsections \ref{NEs} and \ref{PSE}} can be computed by a parametric bootstrap procedure \KLB{(also known as quasi-Bayesian Monte Carlo approach)}. $R$ random vectors of \KLB{the working model} parameters $\bfs{\theta}^r$ ($r=1,...,R$) are drawn from the asymptotic distribution $\mathcal{N }(\bfs{\hat{\theta}},\widehat{V(\bfs{\hat{\theta})}})$ where $\bfs{\hat{\theta}}$ and $\widehat{V(\bfs{\hat{\theta})}}$ are the maximum likelihood estimates and their {Hessian-based variance estimate}. The causal effects are computed for each draw $r$ and the 95\% confidence interval is given by their 2.5\% and 97.5\% percentiles.

\section{Numerical evaluation by simulations}

\KLB{We conducted a series of simulation studies to assess the properties and behavior of our methodology without and with time-varying confounders, under two potential timescales: time in study (Scenarios 1), and age (Scenarios 2).}
 
\subsection{Scenarios 1 according to time in the study}

\subsubsection{Data generating mechanism}\label{datageneration}

For each individual $i$ in a sample of size $N$, we generated an exposure variable {$X_i$} according to a Bernoulli distribution (with probability 0.6). Depending on the presence or absence of a time-varying confounder $\mathcal{L}_t$, we considered a system of two or three processes using the working model defined in SECTION \ref{working_m}. We assumed a constant rate of change for all the processes with random intercepts and simple effects of the covariate on both the initial level and instantaneous change of each process. We also assumed, as in Figure \ref{tab:Fig1}$B$, that only process $\mathcal{L}_t$ impacted the change in $\mathcal{M}_t$, and that processes $\mathcal{L}_t$ and $\mathcal{M}_t$ impacted the change in $\mathcal{Y}_t$ \KLB{(see Web Supplementary Material SECTION 2 for details on the model specification and parameter values). A discretization step of 0.1 was used in the model. Irregular timings of error-prone observations were generated by adding subject-and-visit specific Gaussian deviations (sd=0.05) around annual planned visits over 5 years. The observation process was truncated using an independent uniform dropout in the follow-up range. We considered samples of 500, and generated 250 samples each time. } 

\subsubsection{Variations of scenario 1}
The main scenario, called 1A, was run without (1A-$\bar{TVC}$) and with (1A-TVC) time-varying confounder. Variations of the scenario in the presence of the time-varying confounder were also investigated (see Supplementary Table S2) by changing the discretization step to 0.05 (1B), the sample size to 250 participants (1C), considering only a sub-set of visits for the intermediate processes (1D), and assuming no dropout (1E).

\subsubsection{Estimands}
The estimands were the direct exposure effect on the outcome \KLB{at time $t$ ($t \in \mathbb{R}^+$)} not via the mediator {nor via the time-varying confounder eventually} (i.e., $NDE$ and $PSE_{XY}$ in equations \eqref{NDE} and \eqref{PSE1}), the indirect exposure effect on the outcome \KLB{at time $t$} via the mediator only (i.e., $NIE$ and $PSE_{XMY}$ {in equations \eqref{NIE}, and \eqref{PSE2})} and, {in presence of time-varying confounder,} the indirect effect of the exposure on the outcome via all the paths through the time-varying confounder (i.e., $PSE_{XL(M)Y}$ {in equation \eqref{PSE3}}). 

\subsubsection{{True generated contrast values}}
\KLB{The true effects were empirically computed by generating all the counterfactual outcomes in a large population. 
%\KLB{We generated multiple datasets using the same seed, assigning different combinations of values (0 or 1) to the variable X. We then compared the resulting outcome values to compute the causal effects.} \CPLcom{this was added for the revision to address a comment? This is more or less what is said below. Probably merge both}. 
The counterfactual outcomes were simulated from the same working multivariate mixed model as in SECTION \ref{datageneration} under the following scenarios of exposure combinations:
\begin{itemize}
    \item Natural effect: value of $Y(t)$ when $X = x$ and $\mathcal{M}_t = \mathcal{M}_t(X=x')$
with $x$ and $x' \in \{0,1\}$
\item Path-specific effect: value of $Y(t)$ when $X = x$, $\mathcal{L}_t = \mathcal{L}_t(X=x')$, and $\mathcal{M}_t = \mathcal{M}_t(X=x'', \mathcal{L}_t)$, with $x$, $x'$ and $x'' \in \{0,1\}$
\end{itemize}
The true effects were the empirical versions of equations \eqref{NIE} and \eqref{NDE}, or \eqref{PSE1}, \eqref{PSE2} and \eqref{PSE3}.}

\subsubsection{Working model}
{In all the scenarios, the working model was the} multivariate mixed model detailed in SECTION \ref{working_m} \KLB{and used for generating the data.} 
%This model is useful for calculating the joint distribution of both the mediator and the time-varying confounder, essential for computing the estimates through Monte Carlo approximation. (c.f. Section \ref{monte_carlo}).  

\subsubsection{Performance measures}
We computed the causal effects at 1, 2, 3, 4 and 5 years. The quality of inference was assessed by the distribution of the relative bias (i.e., the bias standardized by the true value (in \%), and the coverage rate of the 95\% confidence interval. 

\subsubsection{Scenario 1 results}
\begin{figure}[htbp]
    \centering
    % Première sous-figure
    \begin{subfigure}[b]{0.45\textwidth}
        \centering
        \includegraphics[width=\textwidth]{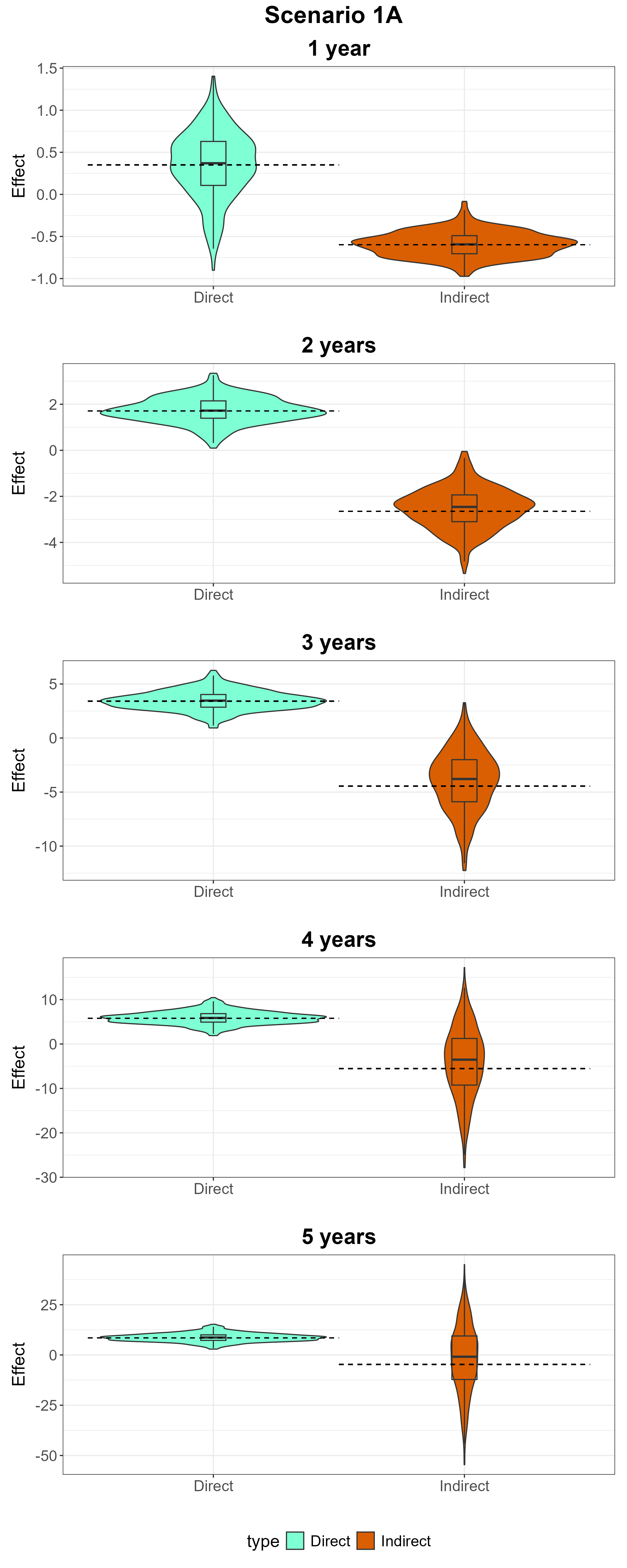} % Remplacez par votre fichier image
        \caption*{a. Without time-varying confounder}
    \end{subfigure}
    \hfill
    % Deuxième sous-figure
    \begin{subfigure}[b]{0.45\textwidth}
        \centering
        \includegraphics[width=\textwidth]{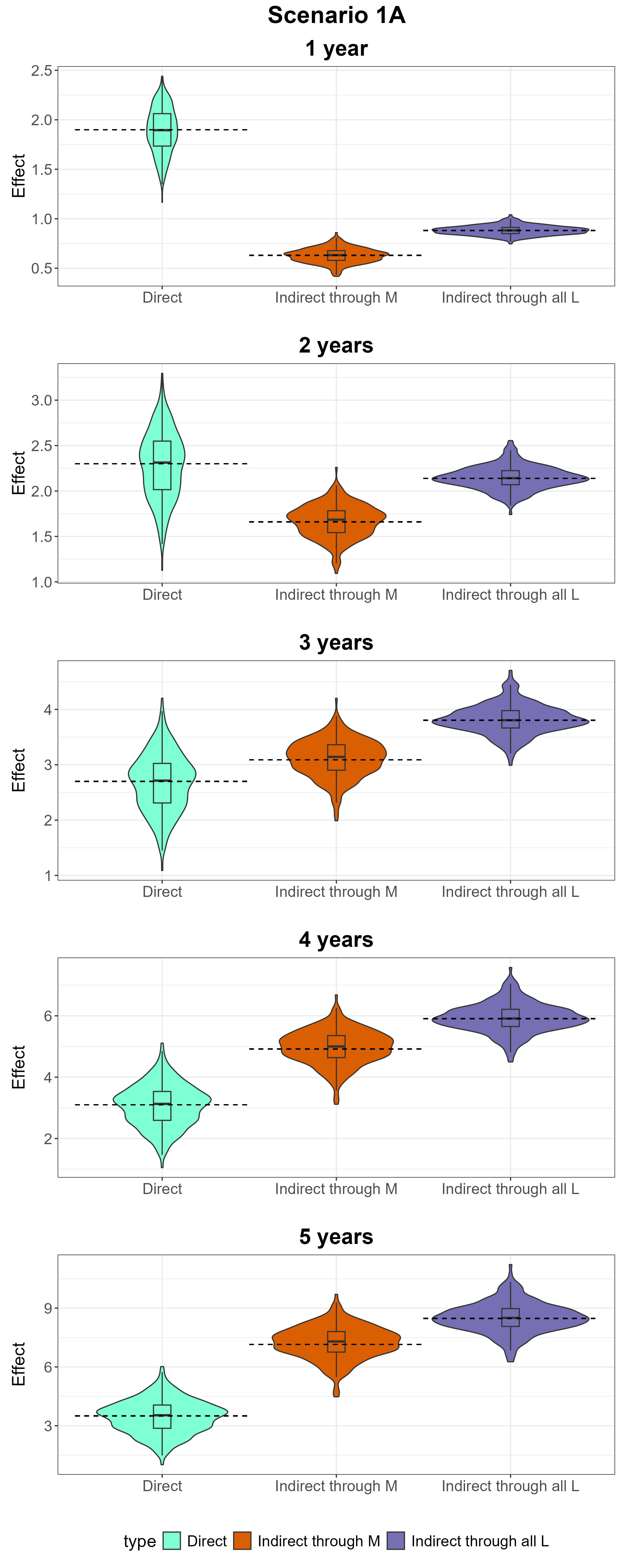} % Remplacez par votre fichier image
        \caption*{b. With time-varying confounder}
    \end{subfigure}
    \caption{Violin plot across 250 Replicates of the estimated causal effects for Scenario 1A without time-varying confounders ((a) direct effect, (b) indirect effect) and with time-varying confounders ((c) direct effect, (d) indirect effect through $\mathcal{M}$, and (e) indirect effect through $\mathcal{L}$ and $\mathcal{M}$).The dashed line represents the true causal effect.}
    \label{figure2}
\end{figure}

For each scenario, the mean relative bias and the coverage rate of the 95\% confidence interval are reported in Table \ref{covR}. Figure \ref{figure2} illustrates the correct estimation of the effects at all times under scenario 1A, without and with time-varying confounder. \KLB{See Web supplementary Figures S1 and S2 for the other scenarios}.
In all the scenarios, the estimates showed minimal bias and coverage rates close to the 95\% nominal value.

\subsection{Scenarios 2 according to age}

\KLB{In a second series of simulations, we mimicked the application setting by considering the subjects entered the cohort at different ages, and age was the time-scale of interest. We also investigated several dropout mechanisms and the situation where intermediate markers have systematic missing visits by design, as encountered with MRI-derived biomarkers in the application. The simulation framework was exactly the same as for scenarios 1, except for the data generating mechanism.}  

\subsubsection{Data generating mechanism}

 \KLB{Differences with Scenarios 1 were that (i) participants entered the cohort according to a Uniform distribution between 65 and 75 years old, (ii) that following the 3C design, the visits were planned at 0, 2, 4, 7, 10, 13, 15, 17 and 20 years after entry with actual visits occurring with a Gaussian random deviation (sd=0.5) around those timings, and that (iii) the generating model included a time-invariant confounder $C$ defined according to a Bernoulli distribution (probability 0.4). Generating model and parameters are given in Supplementary SECTION 2 and the discretization step was 1 year.}
 
\subsubsection{Variations of scenario 2}

\KLB{Variations of this main scenario 2, called 2A, were also investigated. In scenario 2B, the mediator and confounder were collected only at 0, 4, and 10 years after entry as for MRI-derived biomarkers in the 3C cohort. In scenarios 2B-2D, we considered alternative dropout mechanisms compared to the independent dropout of scenario 2A. In 2B, there was no dropout. In 2C, we generated a MAR mechanism by stopping the observation process after the outcome was observed above a threshold. In 2D, we generated a Missing Not At Random (MNAR) mechanism by stopping the observation process exactly when outcome would have been observed above a threshold. In both 2C and 2D, the threshold was chosen at 75\% of the outcome distribution. }
 
\subsubsection{Scenario 2 results}
\KLB{As for scenario 1, when considering cohort entry at heterogeneous ages and age as the time-scale, the path-specific estimates were correctly estimated with negligible bias and coverage rates close to the 95\% nominal value (Table \ref{covR}, Supplementary Figures S3 and S4). This includes case 2B where the mediator and the confounder are measured at most three times compared to the outcome which is measured up to 7 times. In line with the properties of the mixed modeling theory, considering no dropout, MCAR and MAR dropout mechanisms led to correct inference. In the specific MNAR dropout case we investigated, 
the path-specific effects showed some bias (up to approximately 20\%). 
%a bias was observed for the indirect effect through \mathcal{M} and  \mathcal{L}. 
The coverage rate of the 95\% confidence interval remained however close to the nominal value. We anticipate that, although the dropout depended on the unobserved outcome value, the accumulated information on the processes over time made the assumption close to MAR.}

\begin{table}
    \centering
    \small
    \begin{tabular}{llccccccccccc}
    \hline
    &&&&& \multicolumn{2}{c}{\textbf{Direct effect}} & \multicolumn{2}{c}{\textbf{Indirect effect}} & \multicolumn{2}{c}{\textbf{Indirect effect}} \\
    && & &&&& \multicolumn{2}{c}{\textbf{ through $\mathcal{M}$}} & \multicolumn{2}{c}{\textbf{ through $\mathcal{M}$ and $\mathcal{L}$}} \\
    & \textbf{Scen.} & \textbf{Dropout} & \textbf{Time} & \textbf{Schedules} & $\overline{RB}$ (\%) & CR (\%) & $\overline{RB}$ (\%) & CR (\%) & $\overline{RB}$ (\%) & CR (\%)  \\
    \hline

    \multirow{10}{*}{\rotatebox{90}{\textbf{delay}}} & 1A & MCAR & 1 & Balanced & -0.6 & 96.2 & 0.0 & 95.6 & 0.2 & 96.8 \\
    & 1A & MCAR & 2 & Balanced & -0.7 & 94.8 & 0.7 & 94.6 & 0.2 & 96.8 \\
    & 1A & MCAR & 3 & Balanced & -0.8 & 95.0 & 1.2 & 93.8 & 0.2 & 96.2 \\
    & 1A & MCAR & 4 & Balanced & -0.8 & 94.2 & 1.5 & 93.4 & 0.2 & 94.6 \\
    & 1A & MCAR & 5 & Balanced & -0.9 & 93.8 & 1.7 & 93.6 & 0.3 & 94.4 \\
    & 1B & MCAR & 1 & Unbalanced & -0.5 & 95.8 & -0.2 & 93.6 & 0.2 & 95.8 \\
    & 1B & MCAR & 2 & Unbalanced & -0.7 & 94.8 & 0.4 & 95.2 & 0.3 & 95.0 \\
    & 1B & MCAR & 3 & Unbalanced & -0.8 & 94.4 & 0.8 & 95.0 & 0.4 & 94.4 \\
    & 1B & MCAR & 4 & Unbalanced & -0.8 & 94.6 & 1.1 & 95.0 & 0.5 & 94.6 \\
    & 1B & MCAR & 5 & Unbalanced & -0.9 & 94.8 & 1.3 & 94.8 & 0.5 & 94.6 \\
    & 1C & MCAR & 1 & Balanced & -0.4 & 96.4 & 0.2 & 94.2 & 0.1 & 96.0 \\
    & 1C & MCAR & 2 & Balanced & -0.4 & 97.2 & 0.5 & 95.2 & 0.2 & 96.2 \\
    & 1C & MCAR & 3 & Balanced & -0.4 & 97.2 & 0.8 & 93.2 & 0.2 & 95.6 \\
    & 1C & MCAR & 4 & Balanced & -0.4 & 96.8 & 0.9 & 93.4 & 0.3 & 95.2 \\
    & 1C & MCAR & 5 & Balanced & -0.4 & 96.8 & 1.0 & 92.2 & 0.3 & 95.4 \\
    & 1D & MCAR & 1 & Balanced & -0.5 & 95.8 & -0.2 & 93.6 & 0.2 & 95.8 \\
    & 1D & MCAR & 2 & Balanced & -0.7 & 94.8 & 0.4 & 95.2 & 0.3 & 95.0 \\
    & 1D & MCAR & 3 & Balanced & -0.8 & 94.4 & 0.8 & 95.0 & 0.4 & 94.4 \\
    & 1D & MCAR & 4 & Balanced & -0.8 & 94.6 & 1.1 & 95.0 & 0.5 & 94.6 \\
    & 1D & MCAR & 5 & Balanced & -0.9 & 94.8 & 1.3 & 94.8 & 0.5 & 94.6 \\
    & 1E & - & 65 & Balanced & -4.4 & 96.6 & 0.0 & 98.2 & 1.2 & 97.4 \\
    & 1E & - & 70 & Balanced & 1.3 & 95.4 & 1.0 & 97.2 & 0.1 & 97.0 \\
    & 1E & - & 75 & Balanced & 1.2 & 96.4 & 3.0 & 98.2 & 0.1 & 96.6 \\
    & 1E & - & 80 & Balanced & 1.5 & 95.6 & 6.4 & 98.0 & 0.7 & 95.4 \\
    & 1E & - & 85 & Balanced & 1.7 & 95.6 & 1.4 & 98.4 & -0.4 & 94.2 \\
    \hline
    \multirow{20}{*}{\rotatebox{90}{\textbf{age}}} & 2A & MCAR & 65 & Balanced & 0.7 & 96.8 & 0.0 & 96.0 & 0 & 98.2 \\
    & 2A & MCAR & 70 & Balanced & 0.1 & 95.2 & 0.1 & 98.2 & -0.1 & 98.2 \\
    & 2A & MCAR & 75 & Balanced & 0.1 & 94.4 & 0.3 & 97.2 & -0.8 & 95.8 \\
    & 2A & MCAR & 80 & Balanced & 0.2 & 94.0 & 1.0 & 97.4 & 3.0 & 95.4 \\
    & 2A & MCAR & 85 & Balanced & 0.3 & 94.6 & 2.3 & 96.6 & -8.6 & 95.2 \\
    & 2B & MCAR & 65 & Unbalanced & 0.1 & 96.6 & 0.1 & 98.2 & 0.1 & 97.4 \\
    & 2B & MCAR & 70 & Unbalanced & 0.1 & 95.4 & 0.5 & 97.2 & -0.2 & 97.0 \\
    & 2B & MCAR & 75 & Unbalanced & 0.1 & 96.4 & 0.8 & 98.2 & 1.0 & 96.6 \\
    & 2B & MCAR & 80 & Unbalanced & 0.1 & 95.6 & 1.4 & 98.0 & -3.4 & 95.4 \\
    & 2B & MCAR & 85 & Unbalanced & 0.2 & 95.6 & 2.9 & 98.4 & -9.0 & 94.2 \\
    & 2C & MAR & 65 & Balanced & 5.9 & 95.2 & -3.6 & 94.2 & 0.2 & 94.2 \\
    & 2C & MAR & 70 & Balanced & 1.0 & 94.8 & -2.6 & 94.0 & -0.6 & 95.0 \\
    & 2C & MAR & 75 & Balanced & 0.4 & 94.0 & -1.8 & 95.8 & -0.4 & 95.0 \\
    & 2C & MAR & 80 & Balanced & 0.2 & 94.0 & -0.2 & 95.8 & 0.5 & 95.4 \\
    & 2C & MAR & 85 & Balanced & 0.1 & 94.4 & 3.1 & 95.8 & 13.4 & 96.0 \\
    & 2D & MNAR & 65 & Balanced & 7.5 & 98.2 & 1.07 & 97.0 & 0.6 & 95.0 \\
    & 2D & MNAR & 70 & Balanced & 1.1 & 98.0 & 3.2 & 96.8 & -0.5 & 95.6 \\
    & 2D & MNAR & 75 & Balanced & 0.3 & 95.8 & 6.9 & 95.4 & -0.9 & 95.2 \\
    & 2D & MNAR & 80 & Balanced & 0.05 & 95.4 & 12.3 & 94.8 & -1.7 & 95.2 \\
    & 2D & MNAR & 85 & Balanced & -0.08 & 95.6 & 21.1 & 94.8 & 15.9 & 95.2 \\
    \hline
    \end{tabular}
    \vspace{0.5cm}
    \caption{Summary of natural effects (scenario 1A -$\overline{TVC}$) and path-specific effects (1A -$TVC$ to 2 -$TVC$) contrasts estimated in the simulation studies. The mean relative bias ($\overline{RB}$ in \%) and the coverage rate (CR in \%) of the 95\% confidence interval are reported.}
    \label{covR}
\end{table}

\section{Application to cerebral aging}

%We applied the methodology to explore the mechanisms of cognitive aging through two examples
%We applied the methodology to investigate the underlying mechanisms of cognitive aging through two examples:

\KLB{We applied the methodology to study cognitive aging mechanisms through two examples:}
\begin{itemize}
    \item[] {\bf Study 1:} We assessed the impact of the educational level on functional dependency investigating the pathways through verbal fluency and depressive symptomatology.
    \item[] {\bf Study 2:} We investigated the influence of cerebral vascular lesions in the relationship between the main genetic factor of dementia, Apolipoprotein E4 gene (ApoE4), and cognitive functioning, accounting for the potential confounding due to neurodegeneration. 
\end{itemize}

%of education on Instrumental Activities of Daily Living (IADL), by distinguishing the portion influenced by depression from that influenced by cognition (\textbf{Study 1}). Then, we investigate the fundamental causal mechanism underlying between the Apolipoprotein E4 gene (APOE4) and cognitive decline. Specifically, our focus is on understanding the influence of vascular lesions in the relationship between APOE4 and cognition, especially in the presence of neurodegeneration, and to quantify what's the part of the effect of neurodegeneration or vascular lesion in the relationship between APOE4 and cognition (\textbf{Study 2}).

\subsection{The Three-City study sample}

\subsubsection{The cohort}
We leveraged the data from the Three-City (3C) study, a prospective population-based cohort designed to investigate the association between vascular diseases and dementia in the elderly \citep{3c_study_group_vascular_2003}. Individuals aged 65 years and older were randomly enrolled in 1999 from the electoral lists of three French cities (Bordeaux, Dijon and Montpellier). A total of 9294 participants underwent a comprehensive health examination and risk factor assessment at baseline, and at follow-up visits every 2-3 years for a duration of up to 20 years. A Magnetic Resonance Imaging (MRI) assessment was also performed on a subsample at baseline, 4 years and only for Bordeaux center at 10 years of follow-up. The analytical sample comprised the 2,213 participants from Bordeaux and Dijon who were genotyped using genome-wide genotyping arrays, underwent at least one MRI scan, had at least one measure for each marker considered in both studies, no missing covariate for exposures or potential confounders and were free of dementia at baseline. 

\subsubsection{Variables of interest}

\begin{figure}[h!]
    \begin{subfigure}{0.7\linewidth}
        \begin{center}
        \caption*{A. Education level on functional dependency, mediated by cognition and depression}
        \includegraphics[width=\linewidth]{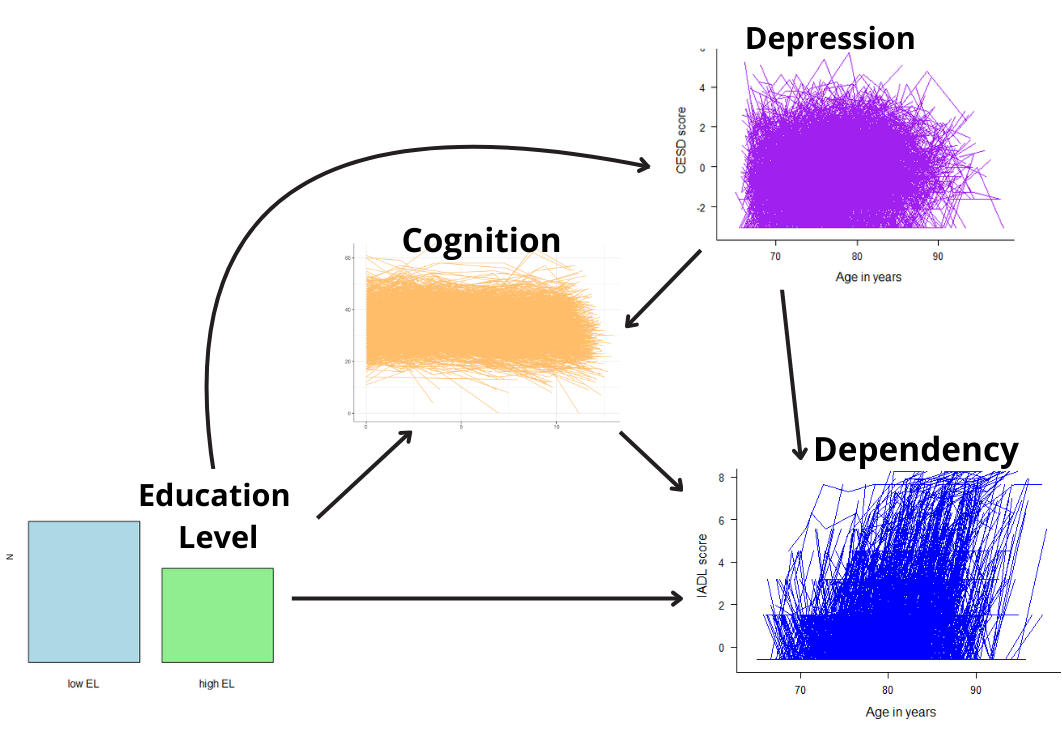}
    \end{center}
        
    \end{subfigure}
    \begin{subfigure}{0.7\linewidth}
        \begin{center}
        \caption*{B. ApoE4 on cognition, mediated by vascular brain lesions and neuro-degeneration }
        \includegraphics[width=\linewidth]{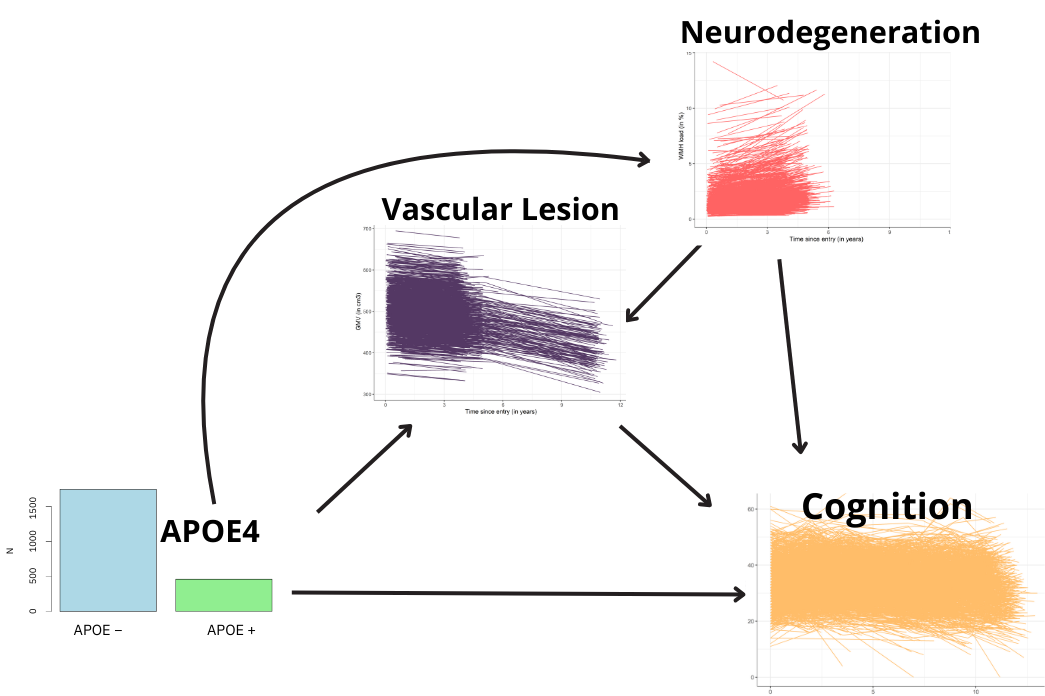}
            \end{center}
    \end{subfigure}
    \caption{{Local Dependence Graph} of Studies 1 and 2; (A) Education level on functional dependency, mediated by cognition and depression (B) ApoE4 on cognition, mediated by vascular brain lesions and global neuro-degeneration}

    \label{fig:Appli_DAG}
\end{figure}

In Study 1 (Figure \ref{fig:Appli_DAG}, {top} panel), the exposure was the binary educational level (high school and higher \textit{versus} lower in reference). The final outcome was the functional dependency measured by {the sum-score} of impairment (the higher, the more dependent) at 5 Instrumental Activities of Daily Living (IADL) (Using {the phone}, transportation, medication management, finances and shopping). The mediator was the verbal fluency measured by the Isaacs Set Test (IST). The IST score equals the count of words provided across four semantic categories within a 15-second interval each. The potential confounder was the depressive symptomatology measured by the score at the Center for Epidemiologic Studies Depression Scale (CESD). The three processes were evaluated at each follow-up, although some \KLB{process-specific missing data arised. By leveraging the data of multiple markers simultaneously, we assumed the MAR mechanism was plausible}. 

In Study 2 (Figure \ref{fig:Appli_DAG}, {bottom} panel), the exposure was the carriership of the ApoE4 and the final outcome was the verbal fluency measured by the IST score. The mediator under investigation was the vascular cerebral lesions measured by the global volume of White Matter Hyperintensities (WMH), and the potential confounder was the global neurodegeneration as measured by the total volume of grey matter (GM). WMH and GM were only collected twice (at 1 and 4 years) in Dijon, and three times (at 1, 4 and 10 years) in Bordeaux Center. In both studies, potential confounders at baseline were sex, age and center (Bordeaux/Dijon).

%We are interested in various variables, such as apolipoprotein E4 gene (APOE4) or education level (EL) measured at baseline. Some data were collected every two to three years during the follow-up, such as cognitive data using the Isaacs Set Test (IST), a measure of verbal fluency that has demonstrated the ability to differentiate early in the pathological progression towards dementia. The score represents the cumulative count of words provided across four semantic categories within a 15-second interval. 

\subsubsection{Description of the analytical sample}
Among the 2,213 participants, 1375 (63.6\%)  were  women,  886 (40.1\%) went to secondary school or higher, and 417 (20.9\%) were APOE4 carriers (Table \ref{AppliDesc}). Participants were 72.3 years old at baseline on average. \KLB{They were followed up for 9.5 years on average with a number of visits per subject ranging from 1 to 8,} a mean of 1.7 (sd=0.7) repeated measures of WMH and GM, of 6 (sd=2.2) repeated measures of IST, and of 5 (sd=1.6) repeated measures of CESD and IADL.

\begin{sidewaystable}[h!]
    \caption{Characteristics of the 2,213 participants of 3C sample according to their APOE4 status, educational level and overall at baseline 
    }
    \tabcolsep=0.02cm
    
\begin{tabular}{llllllllllll}

 \hline 
\multicolumn{2}{c}{Characteristics} & \multicolumn{2}{c}{APOE4 (N=417)} & \multicolumn{2}{c}{No APOE4 (N=1583)} & \multicolumn{2}{c}{High education (N=886)} & \multicolumn{2}{c}{Low education (N=1327)} & \multicolumn{2}{c}{ Overall (N = 2213)} 
\\ 
&& N (\%) & Mean (SD) & N (\%) & Mean (SD) & N (\%) & Mean (SD)  & N (\%) & Mean (SD) & N (\%) & Mean (SD)
\\
\hline
\multicolumn{2}{l}{Sex} &&&&&&&&&& \\
&\textit{female} & 60.1 && 62.7 && 54.3 && 67.4 && 62.1 & \\
&\textit{male} & 39.9 && 37.3 && 45.7 && 32.6 && 36.4 & \\
\multicolumn{2}{l}{Age at entry} && 71.9 (3.8) && 72.1 (4.0) && 72.3 (4.0)&& 72.0 (4.0) && 72.0 (4.0) \\
\multicolumn{2}{l}{WMH ($cm^3$)} & & 2.1 (1.6) && 2.3 (2.2) && 2.3 (2.2) && 2.3 (2.0) && 2.3 (2.1) \\
\multicolumn{2}{l}{Number of WMH } &&1.5 (0.6) && 1.5 (0.6) && 1.6 (0.6)&& 1.5 (0.6)&& 1.5 (0.6)\\
\multicolumn{2}{l}{measures/subject} && &&  && && &&\\

\multicolumn{2}{l}{GM ($cm^3$)} && 502.6 (51.2) && 496.3 (49.1) && 507.5 (49) && 491.4 (49) && 497.6 (49.6)\\
\multicolumn{2}{l}{Number of GM } && 1.6 (0.7) && 1.6 (0.7) && 1.7 (0.8) && 1.6 (0.7)&& 1.6 (0.7)\\
\multicolumn{2}{l}{measures/subject} && &&  &&\\

\multicolumn{2}{l}{IST score [0-66]} & & 32.9 (7.1) && 33.5 (6.7) && 35.0 (6.8) && 32.2 (6.5) && 33.4 (6.8)\\
\multicolumn{2}{l}{Number of IST } && 6.4 (2.3) && 6.5 (2.1) && 6.7 (2.1)&& 6.4 (2.2)&& 6.5 (2.2)\\
\multicolumn{2}{l}{measures/subject} && &&  &&\\

\multicolumn{2}{l}{CESD score [0-60]} & & 9.2 (8.3) && 9.1 (8.3) && 8.3 (7.7) && 9.6 (8.6) && 9.1 (8.2)\\
\multicolumn{2}{l}{Number of CESD } && 4.8 (1.6) && 5.0 (1.5) && 5.1 (1.4)&& 4.8 (1.6)&& 5.0 (1.6)\\
\multicolumn{2}{l}{measures/subject} && &&  &&\\

\multicolumn{2}{l}{IADL score [0-5]} & & 0.5 (1.4) && 0.4 (1.2) && 0.4 (1.2) && 0.5 (1.3) && 0.4 (1.3)\\
\multicolumn{2}{l}{Number of IADL } && 5.1 (1.6) && 5.2 (1.6) && 5.3 (1.4)&& 5.0 (1.6)&& 5.1 (1.6)\\
\multicolumn{2}{l}{measures/subject} && &&  &&\\

\multicolumn{2}{l}{Years of follow-up}&&9.1 (4.2) && 9.5 (3.9)  &&9.8 (3.9) &&9.2 (4.0) &&9.5 (4.0) \\
\hline
\multicolumn{12}{l}{Abbreviations: N=sample size, WMH=White matter hyper-intensities, GM=grey matter, IST=Isaacs Set Test,}\\
\multicolumn{12}{l}{CESD = Center for Epidemiologic Studies Depression Scale (normalized score),} \\
\multicolumn{12}{l}{IADL = Instrumental Activities of Daily Living (normalized score), SD=standard deviation}
\end{tabular}\label{AppliDesc}
\end{sidewaystable}

\subsection{Path-specific effects}

\subsubsection{Working model specification and estimands}
Path-specific contrasts were estimated using the working model defined in Equations \eqref{structworkmod} and \eqref{obsworkmod}. In both studies, the timescale was the age and the discretization step was of 1 year. The {working} model assumed a constant change over age. For each component, both the initial level and the change over time were adjusted for ApoE4, educational level, age at entry in the cohort, sex and center, and included an individual random-effect. In study 2, the Grey matter volume model was further adjusted for the total intra-cranial volume. In both studies, the effect of $\mathcal{L}$ on $\mathcal{M}$, and the effect of $\mathcal{L}$ and $\mathcal{M}$ on $\mathcal{Y}$ were in interaction with the exposure variable. {The specifications of the two working models are detailed in Web Supplementary SECTION 3.}

To satisfy the model's assumptions, WMH volume was log transformed, and CESD and IADL scores were normalized in a preliminary step using quadratic I-splines \citep{proust-lima_analysis_2013}. In the following, IADL normalized score measuring functional dependency is expressed in Standard Deviation of the population at 65 years old. The estimates of the working models for the two studies are reported in Web Supplementary Tables S4 and S5. 

\subsubsection{Identifiability assumptions}
\KLB{Given the likely time-varying confounding of depressive symptomatology in Study 1 and neurodegeneration in Study 2, the sequential ignorability of the relations cognition/function and vascular lesions/cognition were unsuitable. We thus investigated path-specific effects rather than natural effects under the full assumptions listed in Subsection \ref{assump}. We assumed however no other time-dependent confounders.}

%assumed that there was no remaining confounding between the exposure and the outcome variables, between the intermediate variables and the outcomes, and between the exposure and the intermediate variables after adjustment for the considered confounders.} 

\subsubsection{Study 1}
\begin{figure}
 \begin{subfigure}{1\linewidth}
    \caption*{A. Educational level on functional dependency}
    \includegraphics[width=1.1\textwidth, trim=0 2cm 0 2cm, clip]{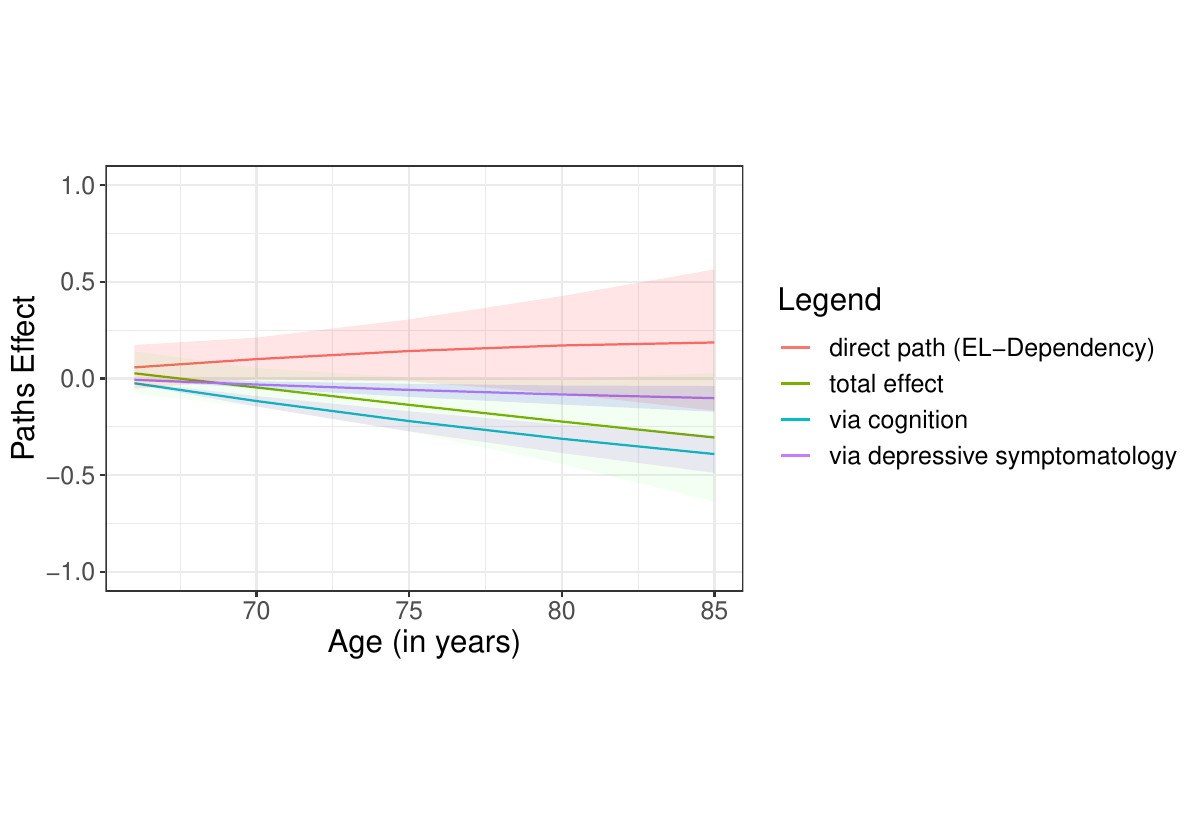}   
\end{subfigure}
    \begin{subfigure}{1\linewidth}
        \caption*{B. ApoE4 on cognitive functioning}
        \includegraphics[width=1.05\textwidth]{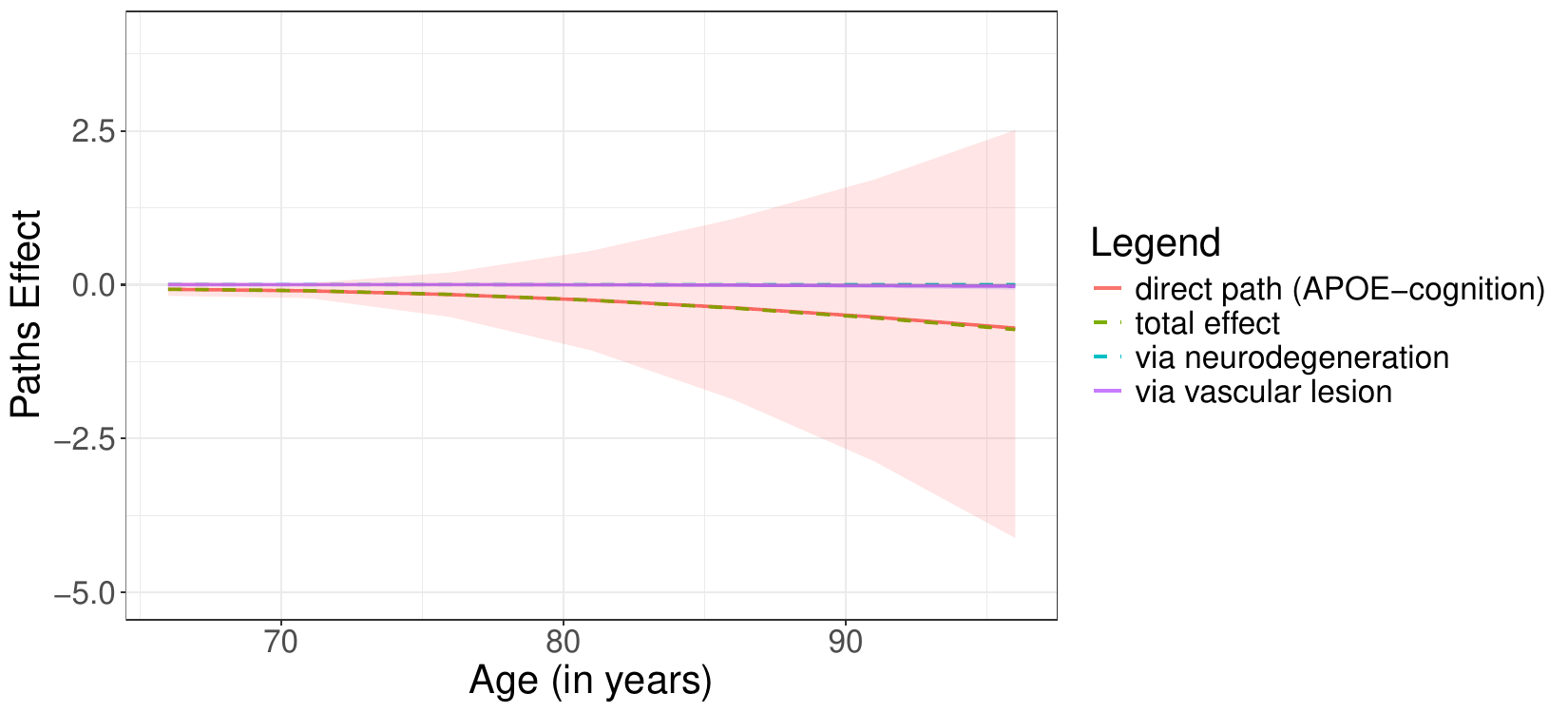}  
    \end{subfigure}
\caption{Estimated path-specific effects in the 3C cohort study: A. high educational level (reference: low) on functional dependency (IADL score) through cognitive functioning (IST score) and depressive symptomatology (CESD score), and B. APOE4 carrier (reference: non carrier) on cognitive functioning (IST score) through neurodegeneration (GM) and vascular lesions( WMH). Confidence bands are obtained by parametric Bootstrap with 1000 replicates. }
    \label{fig:Appli}
\end{figure}

The path-specific effects of educational level on functional dependency are plotted in Figure \ref{fig:Appli} (top panel A). Overall, a higher educational level induced a lower functional impairment at all ages, with a difference increasing with age. This effect was largely mediated by the path through cognitive functioning. It was also slightly due to the path through depressive symptomatology, both in the direction of higher educational level implying lower functional impairment. The direct effect of educational level on functional level was in the opposite direction but not significant. 

\subsubsection{Study 2}
The path-specific effects relating the presence of allele APOE4 %$\epsilon$ 4 of the Apolipoprotein $E$ 
to cognitive functioning are displayed in Figure \ref{fig:Appli} (bottom panel). In this example, the total effect of ApoE4 on cognitive functioning was driven by its direct effect with a systematically lower cognitive level for ApoE4 carriers. The two path-specific effects through the vascular lesions as measured by the White Matter Hyper-intensities, and through the neuro-degeneration confounding factor as measured by the total volume of Grey Matter were negligible.
%{However these results should be interpreted with caution as they are very likely biased. As shown in the simulation scenario 2, in such a context with only a few measures for the intermediate variables (average of 1.6 (sd=0.7) measures per subject), the path-specific effects cannot be correctly retrieved.} 
%We conducted further simulations to assess the influence of the number of measurements and confirmed, as depicted in Figure \ref{PSE:Age}, that biases are indeed present in the estimation of our causal effects. After some investigations, it appears that this stems from biases in the estimates of the working model used, thus impacting the causal effects.

\section{Discussion}

In this paper, we have expanded the methodology of mediation analysis by proposing an approach for mediator, confounder and outcome that are processes defined in continuous time and measured at sparse and possibly irregular visits in prospective studies. In addition to the natural indirect and direct effects that only hold in the absence of time-dependent confounders, we developed the path-specific effects to address mediation analyses in the presence of time-dependent confounders. A simulation study underlined that the causal contrasts, derived from a \KLB{well-specified} multivariate longitudinal working model based on differential equations, were correctly estimated \KLB{under several settings mimicking cohort data including cases in which the intermediate processes were less frequently measured.} 
%Two applications further illustrated the methodology in cerebral aging research. 

%provided the repeated information collected on the processes in play was rich enough. This prevented for instance the application of mediation analysis to assess the mediating effect of a genetic factor on cognition through MRI-derived features that were measured only two or three times in a population-based cohort.} 

%We illustrated our methodology, two studies on brain aging were conducted using data from the 3C cohort. The first study examines functional aging by investigating the impact of education level on functional dependence, considering mediation through cognition and controlling for depression. The second study focuses on the influence of APOE on cognition, considering mediation through vascular lesions and controlling for neurodegeneration.

\KLB{Motivated by pathways in cerebral aging, we primarily focused on natural effects and adapted the 5 required assumptions to make these quantities identifiable with longitudinal data. Given the absence of time-varying counfounders postulated by the classical sequential ignorability assumption is very unlikely in epidemiology, we proposed path-specific effects that allow decomposing the indirect effects by separating the pathways through the mediator alone and through a time-varying confounder. For the estimands identification, we restricted our setting to the case where the history of the outcome doesn't affect later mediator values. This assumption is realistic in the application, considering the history of cognition may not impact future vascular lesions and neurodegeneration, or the functional dependency history may not impact future cognitive level and depressive symptomatology. We leave to future work the relaxing of the sequential ignorability assumption to other situation. Furthermore, we defined marginal effects in the sense that they include the intermediate history of the outcome as part of the pathways of interest. Marginal effects are to be distinguished from conditional effects for which the pathways through the history of Y would be eliminated, and only the effect of the last instant would be sought. In the second application for instance, we were interested in separating the effect of ApoE4 on the cognitive level at a certain age that was due to the effect of ApoE4 on the whole history of vascular lesions up to this age, not only what remained after controlling for the cognitive history up to this age. The indirect effect of the intervention through vascular lesions had to include how these lesions changed the cognitive history over age towards the cognitive level at the age of interest.}

%Ici, c'est une discussion qui est attendue. \KLB{Among assumptions, we used the consistency assumption : ${Y}(t)$=$Y_t(x, m_t) $ if X=x, $\mathcal{M}_t=m_t$; the positivity assumption, it means that $P(X=x|C) > 0$ and {$P(\mathcal{M}_t(x)=m_t|X=x,C) > 0 $}, the sequential ignorability (i.e. $Y_t(x, m_t) \indep X|C$,          $Y_t(x, m_t) \indep \mathcal{M}_t|C,X$,        $\mathcal{M}_t(x) \indep X|C$) and the cross-world independence assumptions $Y_t(x,m_t) \indep \mathcal{M}_t(x')|X,C$}
\KLB{The path-specific effects still rely on the cross-world ignorability assumption. In the case of a too likely violation of this assumption, our methodology can also extend the stochastic intervention approach to mediator, confounder and outcomes defined in continuous time. Stochastic effects can be obtained by replacing the process $\mathcal{M}_t$ in the contrast definitions (e.g., equations \eqref{NE}, \eqref{NIE}, \eqref{NDE} for natural effects) with a stochastic intervention $G_{\mathcal{M}_t}$} As not requiring the cross-world independence assumption, they apply more broadly notably in the presence of an exposure-induced confounder \citep{vanderweele_effect_2014}. Their interpretation differs from natural effects as in general they measure the impact of interventions at the population level, rather than mediating mechanisms  \citep{moreno-betancur_understanding_2018}. 

%In contrast to natural effects, which are defined based on interventions at the individual level, interventional effects are defined with regard to interventions at the population level (\citep{moreno-betancur_understanding_2018}).

Causal contrasts \KLB{are defined at the level of the latent processes underlying the observations, thus in continuous-time. Their} estimation requires the use of a working model \KLB{for the observations} from which the conditional distributions of mediator and outcome processes can be derived. We used in this study a multivariate mixed-effects model based on differential equations \citep{tadde_dynamic_2020} to estimate the joint distribution of all the processes in play \KLB{following the estimand framework while handling the irregularity in the number and timings of observations across individuals and processes. By specifying that the instantaneous change in the outcome was explained by the current values of the outcome, mediator, and confounder, the working model assumed that the entire history of the mediator and confounder processes were associated with the outcome trajectory. In contrast, although not a requirement of the modeling technique, we assumed the mediator was not affected by the past values of the outcome to align with the sequential ignorability assumption.} \KLB{Alternative working models could be considered to estimate the proposed causal effects provided they can adequately handle the data characteristics, and capture the interrelationships between processes in play.} \KLB{The working model considered here exploits all the available longitudinal information under a MAR mechanism following the mixed model theory in order to retrieve the structure of relationships between the underlying processes which are defined in continuous time. This allows a continuous-time definition of the processes in play despite the collection of observations at irregular and possibly sparse times. The method however requires the observation times to be uninformative, and further assumes independent measurement errors. We leave for future work extensions to handle additional data complexities such as informative censoring or informative observation times as they would necessitate further statistical development \citep{coulombe_estimating_2021,pullenayegum_causal_2023}. The methodology also requires complete cases for the time-invariant confounders and exposure, and at least one observation of each time-varying variable, and it requires the model to be correctly specified.}

In a first application, we found an overall effect of education in functional dependency suggesting that education may promote greater autonomy. This effect, primarily influenced by the mediation through cognition and to a lesser extent through depression, underscores the importance of considering these factors in understanding the dynamics of functional aging. The second application exploring the impact of the main genetic factor of Alzheimer's disease \citep{reitz_epidemiology_2011}, ApoE4, on cognitive level in the elderly suggested that the total effect of ApoE4 on cognitive functioning was not mediated by its effect through vascular lesions or overall grey matter atrophy as a proxy of neurodegeneration. \KLB{We assumed the grey matter volume captured the time-varying confounding by neurodegeneration of the relationship between vascular lesions and cognition. If available, we could have considered additional biomarkers (e.g., Tau protein). The methodology would remain very similar and capture the overall pathway through the multivariate confounding processes altogether.} 
%Deja dit dans l'appli : In both applications, by considering multiple markers simultaneously, we assumed that the MAR assumption became plausible as the observed markers are likely to capture the underlying process leading to censoring.

%However, these results should not be interpreted further. We chose to report them in order to emphasize the limits of mediation analyzes on longitudinal data. Indeed, the simulation study showed that causal contrasts could not be retrieved correctly when the repeated information was poor as in this case with a couple of repeated measures for the intermediate processes. 

%\KLB{The results we obtain in our applications depend on the reliability of our assumptions.. For sequential ignorability, regardless of the application context, if there are confounding factors between X and Y that are omitted, the assumption will be biased. This is why, in our applications, we assume, as in many studies, that there are no unmeasured confounding factors. \\Unlike the two previous assumptions, the positivity assumption is testable based on the data available in our application. To do this, we can use contingency tables. This approach allows us to verify that the probabilities stated in the assumption (ii) are not zero for all combinations of covariate values.}

{Mediation analysis had already been extended to longitudinal data. Yet methods were mainly restricted to discrete time when processes usually lie in continuous time and are measured at irregular timings across individuals, and possibly across variables. By defining the causal contrasts at the process level and using a working model adapted to irregular longitudinal data, our methodology goes one step further to address mediation questions related to time-fixed exposures in prospective cohorts. }

\section*{Funding}
    This work was funded by the French government in the framework of the PIA3 ("Investment for the future") (project 17-EURE-0019) and an EHESP Doctoral Network Fellowship. It was also carried out within the University of Bordeaux's France 2030 program / RRI PHDS, and the DyMES project funded by the French National Research Agency (ANR-18-CE36-0004). 
%The Three-City Study is conducted under a partnership agreement between the Institut National de la Santé et de la Recherche Médicale (INSERM), the Victor Segalen– Bordeaux II University, and Sanofi Aventis. The Fondation pour la Recherche Médicale funded the preparation and initiation of the study. The 3C Study is also supported by the Caisse Nationale Maladie des Travailleurs Salariés, Direction Générale de la Santé, MGEN, Institut de la Longévité, Conseils Régionaux of Aquitaine and Bourgogne, Fondation de France, and Ministry of Research–INSERM.

\section*{Acknowledgments}
The authors would like to thank the investigators of the 3C study for providing access to data from the Bordeaux and Dijon centers.

\KLB{\section*{Supplementary Materials}
Web Appendices, Tables, Figures referenced in Sections 2, 3 and 4 are available with this paper at the Biometrics website on Oxford Academic. 
\section*{Data availabilty}
Scripts for replicating the application and the simulation runs are provided in supplementary material. This replication material does not include the application data. Anonymized data can only be shared by reasonable motivated request to the 3C scientific committee.
}
\newpage
%\bibliographystyle{bbl}
%\bibliographystyle{unsrt}
%\bibliography{med}

\begin{thebibliography}{}

\bibitem[\protect\citeauthoryear{{3C Study Group}}{{3C Study
  Group}}{2003}]{3c_study_group_vascular_2003}
{3C Study Group} (2003).
\newblock Vascular factors and risk of dementia: design of the {Three}-{City}
  {Study} and baseline characteristics of the study population.
\newblock {\em Neuroepidemiology} {\bf 22,} 316--325.

\bibitem[\protect\citeauthoryear{Albert, Li, Sun, Woyczynski, and
  Nelson}{Albert et~al.}{2019}]{albert_continuous-time_2019}
Albert, J.~M., Li, Y., Sun, J., Woyczynski, W.~A., and Nelson, S. (2019).
\newblock Continuous-time causal mediation analysis.
\newblock {\em Statistics in Medicine} {\bf 38,} 4334--4347.

\bibitem[\protect\citeauthoryear{Avin, Shpitser, and Pearl}{Avin
  et~al.}{2005}]{avin_identifiability_nodate}
Avin, C., Shpitser, I., and Pearl, J. (2005).
\newblock Identifiability of {Path}-{Specific} {Effects}.

\bibitem[\protect\citeauthoryear{Bind, Vanderweele, Coull, and Schwartz}{Bind
  et~al.}{2016}]{bind_causal_2016}
Bind, M.-A.~C., Vanderweele, T.~J., Coull, B.~A., and Schwartz, J.~D. (2016).
\newblock Causal mediation analysis for longitudinal data with exogenous
  exposure.
\newblock {\em Biostatistics} {\bf 17,} 122--134.

\bibitem{proust-lima_analysis_2013}
C.~Proust-Lima, H.~Amieva, and H.~Jacqmin-Gadda, ``Analysis of multivariate mixed longitudinal data: a flexible latent process approach,'' \emph{The British journal of mathematical and statistical psychology}, vol.~66, no.~3, pp. 470--487, Nov. 2013. doi: \url{10.1111/bmsp.12000}.


\bibitem{pullenayegum_causal_2023}
E.~M. Pullenayegum, C.~Birken, J.~Maguire, et~al., ``Causal inference with longitudinal data subject to irregular assessment times,'' \emph{Statistics in Medicine}, vol.~42, no.~14, pp. 2361--2393, 2023. [Online]. Available: \url{https://onlinelibrary.wiley.com/doi/abs/10.1002/sim.9727}

\bibitem{coulombe_estimating_2021}
J.~Coulombe, E.~E.~M. Moodie, and R.~W. Platt, ``Estimating the marginal effect of a continuous exposure on an ordinal outcome using data subject to covariate-driven treatment and visit processes,'' \emph{Statistical Medicine}, vol.~40, no.~26, pp. 5746--5764, 2021. doi: \url{10.1002/sim.9151}



\bibitem[\protect\citeauthoryear{Booth and Sarkar}{Booth and
  Sarkar}{1998}]{booth_monte_1998}
Booth, J.~G. and Sarkar, S. (1998).
\newblock Monte {Carlo} {Approximation} of {Bootstrap} {Variances}.
\newblock {\em The American Statistician} {\bf 52,} 354--357.

\bibitem[\protect\citeauthoryear{Huang and Yang}{Huang and
  Yang}{2017}]{huang_causal_2017}
Huang, Y.-T. and Yang, H.-I. (2017).
\newblock Causal mediation analysis of survival outcome with multiple
  mediators.
\newblock {\em Epidemiology} {\bf 28,} 370--378.

\bibitem[\protect\citeauthoryear{Laird and Ware}{Laird and
  Ware}{1982}]{laird_random-effects_1982}
Laird, N.~M. and Ware, J.~H. (1982).
\newblock Random-effects models for longitudinal data.
\newblock {\em Biometrics} {\bf 38,} 963--974.

\bibitem[\protect\citeauthoryear{Lange and Hansen}{Lange and
  Hansen}{2011}]{lange_direct_2011}
Lange, T. and Hansen, J.~V. (2011).
\newblock Direct and {Indirect} {Effects} in a {Survival} {Context}.
\newblock {\em Epidemiology} {\bf 22,} 575--581.

\bibitem[\protect\citeauthoryear{Lange, Rasmussen, and Thygesen}{Lange
  et~al.}{2014}]{lange_assessing_2014}
Lange, T., Rasmussen, M., and Thygesen, L.~C. (2014).
\newblock Assessing natural direct and indirect effects through multiple
  pathways.
\newblock {\em Am J Epidemiol} {\bf 179,} 513--518.

\bibitem[\protect\citeauthoryear{Mittinty and Vansteelandt}{Mittinty and
  Vansteelandt}{2020}]{mittinty_longitudinal_2020}
Mittinty, M.~N. and Vansteelandt, S. (2020).
\newblock Longitudinal {Mediation} {Analysis} {Using} {Natural} {Effect}
  {Models}.
\newblock {\em Am J Epidemiol} {\bf 189,} 1427--1435.

\bibitem[\protect\citeauthoryear{Moreno-Betancur and Carlin}{Moreno-Betancur
  and Carlin}{2018}]{moreno-betancur_understanding_2018}
Moreno-Betancur, M. and Carlin, J.~B. (2018).
\newblock Understanding {Interventional} {Effects}: {A} {More} {Natural}
  {Approach} to {Mediation} {Analysis}?
\newblock {\em Epidemiology} {\bf 29,} 614.

\bibitem[\protect\citeauthoryear{Nguyen, Schmid, Ogburn, and Stuart}{Nguyen
  et~al.}{2022}]{nguyen_clarifying_2022}
Nguyen, T.~Q., Schmid, I., Ogburn, E.~L., and Stuart, E.~A. (2022).
\newblock Clarifying causal mediation analysis: {Effect} identification via
  three assumptions and five potential outcomes.
\newblock {\em Journal of Causal Inference} {\bf 10,} 246--279.

\bibitem[\protect\citeauthoryear{Philipps, Hejblum, Prague, Commenges, and
  Proust-Lima}{Philipps et~al.}{2021}]{philipps_robust_2021}
Philipps, V., Hejblum, P., B., Prague, M., Commenges, D., and Proust-Lima, C.
  (2021).
\newblock Robust and {Efficient} {Optimization} {Using} a
  {Marquardt}-{Levenberg} {Algorithm} with {R} {Package} {marqLevAlg}.
\newblock {\em The R Journal} {\bf 13,} 273.

\bibitem[\protect\citeauthoryear{Reitz, Brayne, and Mayeux}{Reitz
  et~al.}{2011}]{reitz_epidemiology_2011}
Reitz, C., Brayne, C., and Mayeux, R. (2011).
\newblock Epidemiology of {Alzheimer} disease.
\newblock {\em Nat Rev Neurol} {\bf 7,} 137--152.

\bibitem[\protect\citeauthoryear{Robins and Greenland}{Robins and
  Greenland}{1992}]{robins_identifiability_1992-1}
Robins, J.~M. and Greenland, S. (1992).
\newblock Identifiability and {Exchangeability} for {Direct} and {Indirect}
  {Effects}.
\newblock {\em Epidemiology} {\bf 3,} 143--155.

\bibitem[\protect\citeauthoryear{Saunders and Blume}{Saunders and
  Blume}{2018}]{saunders_classical_2018}
Saunders, C.~T. and Blume, J.~D. (2018).
\newblock A classical regression framework for mediation analysis: fitting one
  model to estimate mediation effects.
\newblock {\em Biostatistics} {\bf 19,} 514--528.

\bibitem[\protect\citeauthoryear{Taddé, Jacqmin-Gadda, Dartigues, Commenges,
  and Proust-Lima}{Taddé et~al.}{2020}]{tadde_dynamic_2020}
Taddé, B.~O., Jacqmin-Gadda, H., Dartigues, J.-F., Commenges, D., and
  Proust-Lima, C. (2020).
\newblock Dynamic modeling of multivariate dimensions and their temporal
  relationships using latent processes: {Application} to {Alzheimer}'s disease.
\newblock {\em Biometrics} {\bf 76,} 886--899.

\bibitem[\protect\citeauthoryear{Tai, Lin, Chu, Yu, Puhan, and VanderWeele}{Tai
  et~al.}{2023}]{tai_causal_2023}
Tai, A.-S., Lin, S.-H., Chu, Y.-C., Yu, T., Puhan, M.~A., and VanderWeele, T.
  (2023).
\newblock Causal {Mediation} {Analysis} with {Multiple} {Time}-varying
  {Mediators}.
\newblock {\em Epidemiology} {\bf 34,} 8.

\bibitem[\protect\citeauthoryear{Valeri, Proust-Lima, Fan, Chen, and
  Jacqmin-Gadda}{Valeri et~al.}{2023}]{valeri_multistate_2023}
Valeri, L., Proust-Lima, C., Fan, W., Chen, J.~T., and Jacqmin-Gadda, H.
  (2023).
\newblock A multistate approach for the study of interventions on an
  intermediate time-to-event in health disparities research.
\newblock {\em Stat Methods Med Res} {\bf 32,} 1445--1460.

\bibitem[\protect\citeauthoryear{VanderWeele}{VanderWeele}{2011}]{vanderweele_causal_2011}
VanderWeele, T.~J. (2011).
\newblock Causal mediation analysis with survival data.
\newblock {\em Epidemiology} {\bf 22,} 582--585.

\bibitem[\protect\citeauthoryear{VanderWeele and Tchetgen~Tchetgen}{VanderWeele
  and Tchetgen~Tchetgen}{2017}]{vanderweele_mediation_2017}
VanderWeele, T.~J. and Tchetgen~Tchetgen, E.~J. (2017).
\newblock Mediation analysis with time varying exposures and mediators.
\newblock {\em J R Stat Soc Series B Stat Methodol} {\bf 79,} 917--938.

\bibitem[\protect\citeauthoryear{Vanderweele, Vansteelandt, and
  Robins}{Vanderweele et~al.}{2014}]{vanderweele_effect_2014}
Vanderweele, T.~J., Vansteelandt, S., and Robins, J.~M. (2014).
\newblock Effect decomposition in the presence of an exposure-induced
  mediator-outcome confounder.
\newblock {\em Epidemiology} {\bf 25,} 300--306.

\bibitem[\protect\citeauthoryear{Vo, Superchi, Boutron, and Vansteelandt}{Vo
  et~al.}{2020}]{vo_conduct_2020}
Vo, T.-T., Superchi, C., Boutron, I., and Vansteelandt, S. (2020).
\newblock The conduct and reporting of mediation analysis in recently published
  randomized controlled trials: results from a methodological systematic
  review.
\newblock {\em J Clin Epidemiol} {\bf 117,} 78--88.

\bibitem[\protect\citeauthoryear{Zheng and Laan}{Zheng and
  Laan}{2012}]{zheng_causal_2012}
Zheng, W. and Laan, M.~J. (2012).
\newblock Causal {Mediation} in a {Survival} {Setting} with {Time}-{Dependent}
  {Mediators}.

\end{thebibliography}

\end{document}

\section{Identification of path-specific effects}\label{sec:supplementary_section_iden}

The path-specific effects are systematically defined as a comparison of two expectations $\mu = \mathbb{E}({Y}_t(x,\mathcal{L}_t(x'),\mathcal{M}_t(x'',\mathcal{L}_t(x')))|C)$ where $x$, $x'$, and $x''$ can take various values. This expectation can be expressed as a function of the observations, rendering it identifiable, thanks to certain assumptions. As the assumptions vary depending on the paths, we will elaborate on the procedure for the three specific path effects: $PSE_{XY}$, $PSE_{XMY}$, $PSE_{XL(M)Y}$.

For $PSE_{XY}$, we have : 

\begin{equation*}
    \mu = \mathbb{E}(Y_t(\textcolor{red}{x},\mathcal{L}_t(x'),\mathcal{M}_t(x',\mathcal{L}_t(x'))))
\end{equation*}

where $\textcolor{red}{x}$ can takes value $x$ or $x'$ 

First, thank to assumption (iii.a) we can conditioned on X, and the estimand can be developed according to the time-confounder history $\mathcal{L}_t$ and the mediator history $\mathcal{M}_t$ jointly. 
\begin{align*}
    \mu  = &  \int_{l_t,m_t}^{} \mathbb{E}(Y_t(x,l_t,m_t)|C=c,X=x, \mathcal{L}_t(x')=l_t,\mathcal{M}_t(x',l_t)=m_t) \times f_{\mathcal{L}_t(x'),\mathcal{M}_t(x')|(C=c),X=x'}~ d_{l_t,m_t}
\end{align*}

Second, thanks to assumptions (iii.c) and (iv.a), we can remove the conditioning $\mathcal{L}_t=l_t$ and $\mathcal{M}_t(x,l_t
) = m_t$ : 

\begin{align*}
    \mu  = &  \int_{l_t,m_t}^{} \mathbb{E}(Y_t(x,l_t,m_t)|C=c, X=x) \times f_{\mathcal{L}_t(x'),\mathcal{M}_t(x')|(C=c, X=x')}~ d_{l_t,m_t}
\end{align*}

Third, using assumption (iii.b) we can condition on  ($L_t, M_t$). 

\begin{align*}
    \mu  = &  \int_{l_t,m_t}^{} \mathbb{E}(Y_t(x,l_t,m_t)|C=c, X=x, \mathcal{L}_t=l_t,\mathcal{M}_t = m_t) \times f_{\mathcal{L}_t(x'),\mathcal{M}_t(x')|(C=c, X=x')}~ d_{l_t,m_t}
\end{align*}

Finally, with the consistency assumption, we obtain : 
 
  \begin{align*}
    \mu  = &  \int_{l_t,m_t}^{} \mathbb{E}(Y_t|C=c, X=x, \mathcal{L}_t=l_t,\mathcal{M}_t = m_t) \times f_{\mathcal{L}_t,\mathcal{M}_t|(C=c, X=x')}~ d_{l_t,m_t}
\end{align*}

\newpage

For $PSE_{XMY}$, we have : 

\begin{equation*}
    \mu = \mathbb{E}(Y_t(x,\mathcal{L}_t(x'),\mathcal{M}_t(\textcolor{red}{x},\mathcal{L}_t(x'))))
\end{equation*}

where $\textcolor{red}{x}$ can takes value $x$ or $x'$ 

First, the estimand can be developed according to the time-confounder history $\mathcal{L}_t$ and the mediator history $\mathcal{M}_t$.

\begin{align*}
    \mu  =  \int_{l_t}^{} \int_{m_t}^{} &\mathbb{E}(Y_t(x,l_t,m_t)|C=c, \mathcal{L}_t(x')=l_t,\mathcal{M}_t(x,l_t)=m_t)\\ \times & 
    f_{\mathcal{L}_t(x')|C=c}(l_t),f_{\mathcal{M}_t(x,l_t)|(C=c, \mathcal{L}_t(x')=l_t)}(m_t)  ~ d_{m_t} d_{l_t}
\end{align*} 

Second, thanks to assumptions (iii.e) and (iv.b), we can remove the conditioning $\mathcal{L}_t=l_t$ and $\mathcal{M}_t(x,l_t
) = mt$ :

\begin{equation*}
   \mu  =\int_{l_t}^{} \int_{m_t}^{} \mathbb{E}(Y_t(x,l_t,m_t)|C=c) 
    f_{\mathcal{L}_t(x')|C=c}(l_t),f_{\mathcal{M}_t(x,l_t)|(C=c}(m_t) ~ d_{m_t} d_{l_t}
\end{equation*}

Third we add the conditioning on
$X = x$ in the expectation of $Y_t$, $X = x'$
in the density of $\mathcal{L}_t$ and $X = x$ and $\mathcal{L_t}$
in the density of $\mathcal{M}_t$ thanks to assumptions (iii.a), (iii.c)
and (iii.e):

\begin{align*}
   \mu  =  \int_{l_t}^{} \int_{m_t}^{}& \mathbb{E}(Y_t(x,l_t,m_t)|C=c, X=x)  \\ \times&
    f_{\mathcal{L}_t(x')|C=c, X=x'}(l_t),f_{\mathcal{M}_t(x,l_t)|(C=c, X=x, \mathcal{L}_t=l_t)}(m_t) ~ d_{m_t} d_{l_t} 
\end{align*}

Then, we add the conditioning on $\mathcal{L}_t = l_t$ and $\mathcal{M}_t = m_t$
in the expectation of $Y_t$ thanks to assumption
(iii.b)

\begin{align*}
   \mu  =  \int_{l_t}^{} \int_{m_t}^{}& \mathbb{E}(Y_t(x,l_t,m_t)|C=c, X=x, \mathcal{L}_t=l_t, \mathcal{M}_t=m_t) \\ \times &
    f_{\mathcal{L}_t(x')|C=c, X=x'}(l_t),f_{\mathcal{M}_t(x,l_t)|(C=c, X=x, \mathcal{L}_t=l_t)}(m_t) ~ d_{m_t} d_{l_t}  
\end{align*}

By applying consistency assumption $\textit{i}$, we obtain:

\begin{align*}
    \mu = \int_{l_t}^{} \int_{m_t}^{}& \mathbb{E}(Y_t|C=c, X=x, \mathcal{L}_t=l_t, \mathcal{M}_t=m_t) \\
    \times &
    f_{\mathcal{L}_t|C=c, X=x'}(l_t),f_{\mathcal{M}_t|(C=c, X=x, \mathcal{L}_t=l_t)}(m_t) ~ d_{m_t} d_{l_t}  
\end{align*}

For $PSE_{XL(M)Y}$ the identification process is the same.

\newpage

\section{Simulation study}

\subsection{Simulation design} \label{sec:supplementary_section_withoutsimu}

We systematically considered the following generation structural models and observation models with the corresponding parameters given in Table \ref{tab:Scena1} and Table \ref{mimic_app} for scenario 1 and 2, respectively. 

\begin{align}
&\textcolor{red}{\text{For process } \mathcal{L}: \left\{
    \begin{array}{ll}
    \begin{aligned}
        L_i(0)  & = \beta_0^{L}+ \beta_1^{L} X_i^{L(0)} + {u}_i^{L} \\
    \frac{\partial L_i(t)}{\partial t} &= \gamma_0^{L} + \gamma_1^{L}{X}_i^{L(t)} +v_i^{L} \\
    \tilde{L}_{ij} &= L_i(t_{ij}) + \epsilon_{ij}^L \text{ for } j=1,...,n_{i}^L\\
    \end{aligned}
    \end{array}\right.}\\
    &\text{For process } \mathcal{M}: \left \{
    \begin{array}{ll}
    \begin{aligned}
         M_i(0)  & = \beta_0^{M}+ \beta_1^{M} X_i^{M(0)} + {u}_i^{M} \\
    \frac{\partial M_i(t)}{\partial t} &= \gamma_0^{M} + \gamma_1^{M}{X}_i^{M(t)} +v_i^{M}  \textcolor{red}{+ \alpha^{ML}L_i(t)}
    \\  
     \tilde{M}_{ij} &= M_i(t_{ij}) + \epsilon_{ij}^M  \text{ for } j=1,...,n_{i}^M\\ 
    \end{aligned}
    \end{array}\right.    \label{mod_ss}\\
    & \text{For process } \mathcal{Y}: \left\{
    \begin{array}{ll}
    \begin{aligned}
         Y_i(0)  & = \beta_0^{Y}+ \beta_1^{Y} X_i^{Y(0)} + {u}_i^{Y} \\
    \frac{\partial Y_i(t)}{\partial t} &= \gamma_0^{Y} + \gamma_1^{Y}{X}_i^{Y(t)} +v_i^{Y} + \alpha^{YM}M_i(t)  \textcolor{red}{ + \alpha^{YL}L_i(t)}    \\
    \tilde{Y}_{ij} &= Y_i(t_{ij}) + \epsilon_{ij}^Y \text{ for } j=1,...,n_{i}^Y\\
    \end{aligned}
    \end{array}\right.
    \label{mod_ss_2}
\end{align}

The process $\mathcal{L}$ sub-equations and corresponding parameters indicated in red were only considered in scenario 2 and in the sub-scenario 1 with time-dependent confounding factors . \\

\begin{table}[H]
    \centering
        \caption{Parameters of the generation models in scenario 1 in absence and presence of time-varying confounder}
    \begin{tabular}{ccc}
        \hline
         \textbf{Parameters} & With L & Without L
         \\
        $\beta_0^{L}$&0.5&-\\
        $\beta_1^{L}$&1.8&-\\
        $\beta_0^{M}$&0.2&0.5\\
        $\beta_1^{M}$&0.9&1.8\\
        $\beta_0^{Y}$&0.6&0.6\\
        $\beta_1^{Y}$&1.5&1.5\\
        $\gamma_0^{L}$&0.1&-\\
        $\gamma_1^{L}$&0.2&-\\
        $\gamma_0^{M}$&0.2&0.1\\
        $\gamma_1^{M}$&0.8&0.2\\
        $\gamma_0^{Y}$&0.8 &0.8\\
        $\gamma_1^{Y}$&0.4 &0.4\\
        chol1& 1&1 \\
        chol2& 0.2&0.1\\
        chol3& 0.1&-\\
        chol6&-&3\\
        chol7&2&-\\
        chol8&0.1&1\\
        chol10&-&2\\
        chol12&3&-\\
        chol16&1& -\\
        chol19&2& -\\
        chol21&3& -\\
        $\alpha_{ML}$ & 0.3 & - \\
        $\alpha_{YL}$ & 0.4 & - \\
        $\alpha_{YM}$ & 0.5 & 0.4\\
        $\sigma_{L}$ & 0.3 & - \\
        $\sigma_{M}$ & 0.6 & 0.3 \\
        $\sigma_{Y}$ & 0.2 & 0.2\\

        \hline
    \end{tabular}
    \label{tab:Scena1}
\end{table}

In the main scenario 1, labelled 1A, we considered samples of 500 subjects, a discretization step of 0.1 year, and a rate of dropout of 10\%. In additional scenarios 1B to 1D in presence of a time-varying confounder, we changed each parameter one by one as listed in Table \ref{tab:Scena param}. In scenario 2, we considered samples of 500 subjects, a discretization step of 1 year, and a rate of dropout of 10\%.

\begin{table}[H]
   \centering
    \caption{Characteristics of sub-scenarios 1A-1D in presence of time-varying confounder}
    \begin{tabular}{ccccc}
        \hline
        \textbf{Scenario} &  \textbf{$N$} & \textbf{${\bf \delta}$ (in years)}  & \textbf{Dropout rate (\%)}\\
        \hline
         \textbf{1A} & 500 & 0.1 & 10 \\
         \textbf{1B} & 500 & 0.05 & 10 \\
         \textbf{1C} & 250 & 0.1  & 10 \\
         \textbf{1D} & 500 & 0.1  & 20 \\
        \hline
    \end{tabular}
    \label{tab:Scena param}
\end{table}

\subsection{Additional results for Scenario 1} 

\begin{figure}[H]
  \centering
  \begin{minipage}[t]{0.55\textwidth}
    \centering
      \caption*{(A) Direct Effect}
    \includegraphics[width=1\linewidth, height=7cm]{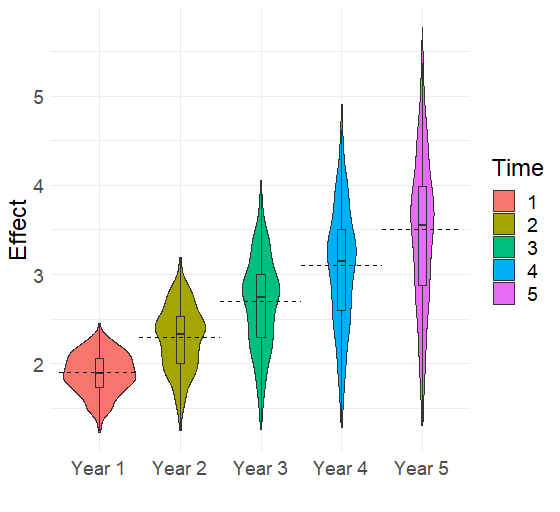} % Remplacez par le chemin de votre première image
  
  \end{minipage}\hfill
  \begin{minipage}[t]{0.55\textwidth}
    \centering
    \caption*{(B) Indirect Effect through $\mathcal{M}$}
    \includegraphics[width=1\linewidth, height=7cm]{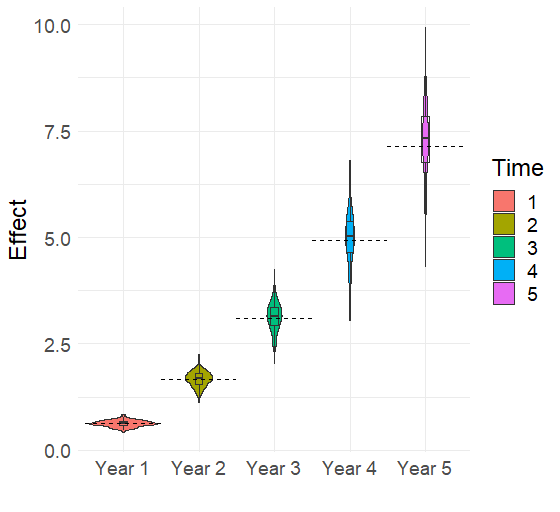} % Remplacez par le chemin de votre deuxième image
    
  \end{minipage}\hfill
  \begin{minipage}[t]{0.55\textwidth}
    \centering
     \caption*{(C) Natural Indirect Effect through $\mathcal{L}$ and $\mathcal{M}$}
    \includegraphics[width=1\linewidth, height=7cm]{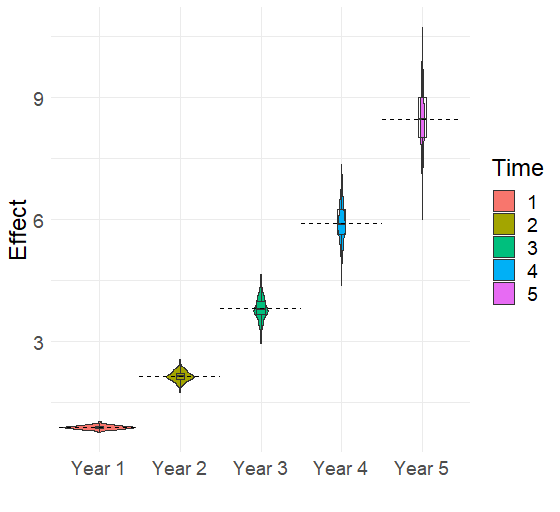} % Remplacez par le chemin de votre deuxième image
   
  \end{minipage}

  \caption{Median Bias of Simulations: Violin Plot Across 250 Replicates for Scenario 1B, with time-varying confounders, for (A) the direct effect, (B) indirect effect through $\mathcal{M}$, and (C) indirect effect through $\mathcal{L}$ and $\mathcal{M}$}
\label{PSE_B}
\end{figure}

\begin{figure}[H]
  \centering
  \begin{minipage}[t]{0.55\textwidth}
    \centering
      \caption*{(A) Direct Effect}
    \includegraphics[width=1\linewidth, height=7cm]{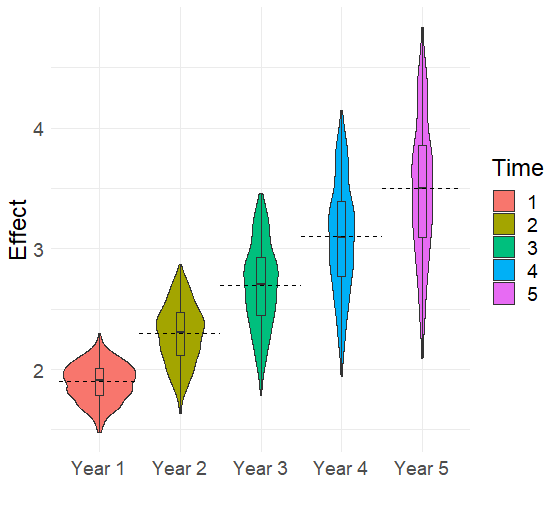} % Remplacez par le chemin de votre première image
  
  \end{minipage}\hfill
  \begin{minipage}[t]{0.55\textwidth}
    \centering
    \caption*{(B) Indirect Effect through $\mathcal{M}$}
    \includegraphics[width=1\linewidth, height=7cm]{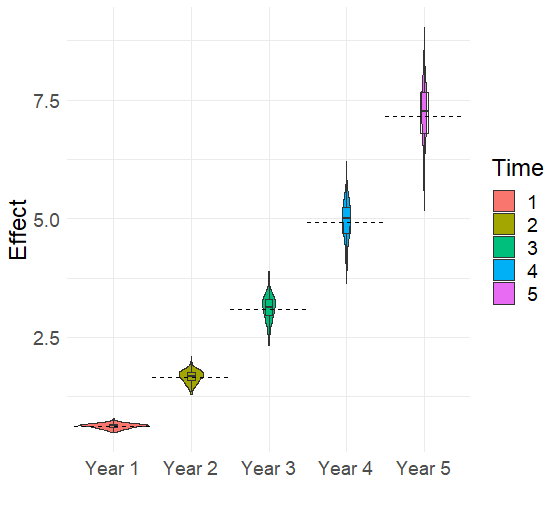} % Remplacez par le chemin de votre deuxième image
    
  \end{minipage}\hfill
  \begin{minipage}[t]{0.55\textwidth}
    \centering
     \caption*{(C) Natural Indirect Effect through $\mathcal{L}$ and $\mathcal{M}$}
    \includegraphics[width=1\linewidth, height=7cm]{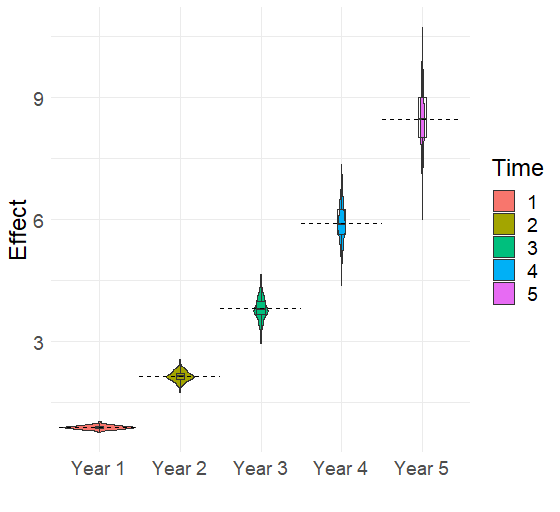} % Remplacez par le chemin de votre deuxième image
   
  \end{minipage}

  \caption{Median Bias of Simulations: Violin Plot Across 250 Replicates for Scenario 1C, with time-varying confounders, for (A) the direct effect, (B) indirect effect through $\mathcal{M}$, and (C) indirect effect through $\mathcal{L}$ and $\mathcal{M}$}
\label{PSE_C}
\end{figure}

\begin{figure}[H]
  \centering
  \begin{minipage}[t]{0.55\textwidth}
    \centering
      \caption*{(A) Direct Effect}
    \includegraphics[width=1\linewidth, height=7cm]{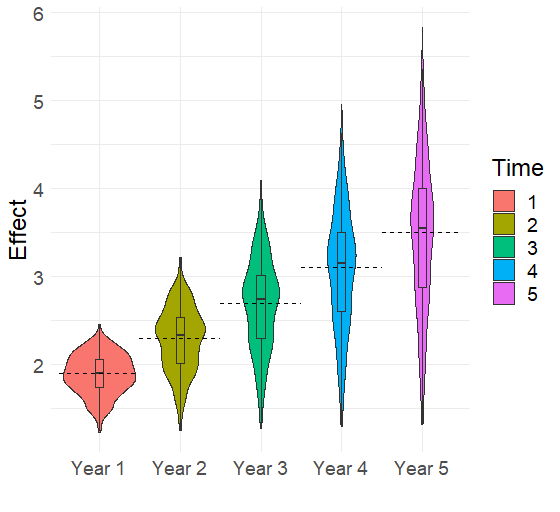} % Remplacez par le chemin de votre première image
  
  \end{minipage}\hfill
  \begin{minipage}[t]{0.55\textwidth}
    \centering
    \caption*{(B) Indirect Effect through $\mathcal{M}$}
    \includegraphics[width=1\linewidth, height=7cm]{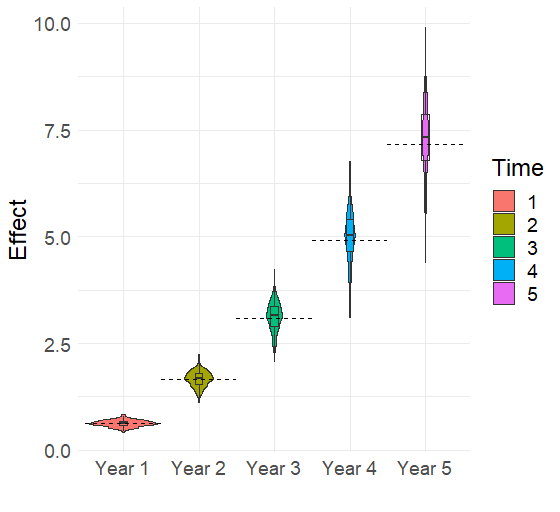} % Remplacez par le chemin de votre deuxième image
    
  \end{minipage}\hfill
  \begin{minipage}[t]{0.55\textwidth}
    \centering
     \caption*{(C) Natural Indirect Effect through $\mathcal{L}$ and $\mathcal{M}$}
    \includegraphics[width=1\linewidth, height=7cm]{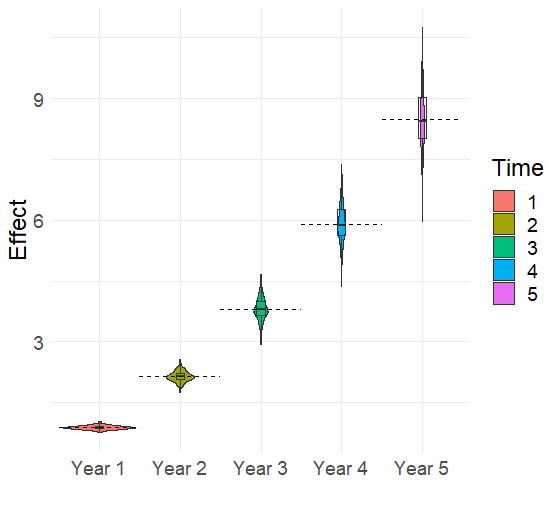} % Remplacez par le chemin de votre deuxième image
   
  \end{minipage}

  \caption{Median Bias of Simulations: Violin Plot Across 250 Replicates for Scenario 1D, with time-varying confounders, for (A) the direct effect, (B) indirect effect through $\mathcal{M}$, and (C) indirect effect through $\mathcal{L}$ and $\mathcal{M}$}
\label{PSE_D}
\end{figure}

The results of scenario 1A are detailed in the main body. Additional results for scenarios B, C, and D in presence of time-varying confounders are given in  Figures \ref{PSE_B},\ref{PSE_C}, and \ref{PSE_D}, respectively. The coverage rate are given in Table 3.

\begin{sidewaystable}[]
    \centering
    \label{covR}
    \small
    \begin{tabular}{cccccccccccccccc}
    \hline
         & \multicolumn{15}{c}{\textbf{Coverage rate (in \%)}} \\
        & \multicolumn{5}{c}{\textbf{Direct effect}} & \multicolumn{5}{c}{\textbf{Indirect effect through M}} & \multicolumn{5}{c}{\textbf{Indirect effect through M and L}}\\
         \textbf{Scenario} &Time1 & Time2 & Time3 & Time4 & Time5& Time1 & Time2 & Time3 & Time4 & Time5& Time1 & Time2 & Time3 & Time4 & Time5 \\
        \hline
        A & 96.2 & 94.8 & 95.0 & 94.2 & 93.8 &95.6 &94.6 &93.8 &93.4 &93.6 &96.8 &96.8 &96.2 & 94.6 & 94.4 \\
        B & 96.0 & 96.4 & 96.2&96.2 & 96.0 & 94.8 &93.2 &93.2 &94.4 &95.6 &95.0 &93.8 &92.4 &92.4 & 92.6 \\ 
        C & 96.4 & 97.2 & 97.2 & 96.8 & 96.8 &94.2 &95.2 &93.2 &93.4 &92.2 &96.0 &96.2 &95.6 & 95.2 & 95.4 \\
        D & 96.6 & 95.2 & 95.0 & 94.6 & 94.6 &95.8 &95.2 &95.2 & 94.4 & 94.4&96.8 &95.8 &95.2 &94.6 &94.2  \\
        Scenario 2 & 30.8 & 14.6 & 11.0 & 10.2 & 10.0 &76.4 &68.6 &63.4 & 55.6 &49.2&1.2 &1.4 &3.0 &3.6 &5.0  \\
        \hline
         \end{tabular}
    \vspace{0.5cm}
    
\caption{Coverage rate in (\%) for the different scenario}
\end{sidewaystable}

\subsection{Additional results for Scenario 2} 

The estimated parameters of the working model are reported in Table \ref{mimic_app}.

\begin{table}[htbp]
\centering
\caption{Estimated parameters of the working model in Scenario 2}
\begin{tabular}{lc}
\hline
\textbf{Parameter}      & \textbf{Value}        \\ \hline
$\beta_0^L$             & -0.20                  \\
$\beta_1^L$             & 0.09                   \\
$\beta_2^L$             & 0.01                   \\
$\beta_0^M$             & 6.12                   \\
$\beta_1^M$             & -0.10                  \\
$\beta_2^M$             & -0.07                  \\
$\beta_0^Y$             & 0.53                   \\
$\beta_1^Y$             & -0.06                  \\
$\beta_2^Y$             & -0.00                  \\
$\gamma_0^L$            & -0.16                  \\
$\gamma_1^L$            & -0.01                  \\
$\gamma_2^L$            & 0.00                   \\
$\gamma_0^M$            & -0.15                  \\
$\gamma_1^M$            & 0.02                   \\
$\gamma_2^M$            & 0.00                   \\
$\gamma_0^Y$            & 0.10                   \\
$\gamma_1^Y$            & -0.01                  \\
$\gamma_2^Y$            & -0.00                  \\
Chol.1                  & 0.86                   \\
Chol.2                  & 0.04                   \\
Chol.3                  & 0.10                   \\
Chol.4                  & -0.00                  \\
Chol.7                  & 1.04                   \\
Chol.8                  & 0.00                   \\
Chol.10                 & 0.06                   \\
Chol.12                 & 0.80                   \\
Chol.15                 & -0.02                  \\
Chol.16                 & 0.03                   \\
Chol.19                 & -0.08                  \\
Chol.21                 & -0.04                  \\
$\alpha_{ML}$.(Intercept) & -0.03                \\
$\alpha_{YL}$.(Intercept) & 0.01                 \\
$\alpha_{YM}$.(Intercept) & -0.01                \\
$\sigma_L$              & 0.22                   \\
$\sigma_M$              & 0.48                   \\
$\sigma_Y$              & -0.51                  \\ \hline
\end{tabular}
\label{mimic_app}
\end{table}

In this scenario with poor repeated information on the mediator and confounder, the working model fails to correctly estimate the parameters which induce bias in the resulting causal contrasts reported in the main document.

\newpage
\section{Additional information about the application}\label{sec:supplementary_section_model}

\subsection{Specification of the working models}

\textbf{Study 1: effect of educational level on the level of functional dependency}
 \begin{align}
&\text{For process } \mathcal{L}: \left\{
    \begin{array}{ll}
    \begin{aligned}
        L_i(0)   = &\beta_0^{L}+ \beta_1^{L} EL_i^{L(0)} + \beta_2^{L}APOE4_i^{L(0)} +
        \beta_3^{L} SEX_i^{L(0)}  +\\
        &
        \beta_4^{L} +CENTRE_i^{L(0)}
        \beta_5^{L}AGE0_i^{L(0)}  +
        {u}_i^{L} \\
    \frac{\partial L_i(t)}{\partial t} =& \gamma_0^{L} + 
    \gamma_1^{L} EL_i^{L} + \gamma_2^{L} APOE4_i^{L} +
        \gamma_3^{L} SEX_i^{L} +\\
        &
        \gamma_4^{L}  CENTRE_i^{L}+
        \gamma_5^{L} AGE0_i^{L} + v_i^{L}
    \end{aligned}
    \end{array}\right.\\
    &\text{For process } \mathcal{M}: \left \{
    \begin{array}{ll}
    \begin{aligned}
         M_i(0)  & = \beta_0^{M}+ \beta_1^{M} EL_i^{M(0)} + \beta_2^{M} APOE4_i^{M(0)} +
        \beta_3^{M} SEX_i^{M(0)} +\\
        &
        \beta_4^{M}  CENTRE_i^{M(0)}+
        \beta_5^{M} AGE0_i^{M(0)} + {u}_i^{M} \\
    \frac{\partial M_i(t)}{\partial t} =& \gamma_0^{M} +  \gamma_1^{M} EL_i^{M} + \gamma_2^{M} APOE4_i^{M} +
        \gamma_3^{M} SEX_i^{M} +\\
        &
        \gamma_4^{M}  CENTRE_i^{M}+
        \gamma_5^{M} AGE0_i^{M} +v_i^{M}  + \alpha^{ML}L_i(t)
    \\  
    \end{aligned}
    \end{array}\right.    \label{mo}\\
    & \text{For process } \mathcal{Y}: \left\{
    \begin{array}{ll}
    \begin{aligned}
          Y_i(0)  & = \beta_0^{Y}+ \beta_1^{Y} EL_i^{Y(0)} + \beta_2^{Y} APOE4_i^{Y(0)} +
        \beta_3^{Y} SEX_i^{Y(0)} +\\
        &
        \beta_4^{Y}  CENTRE_i^{Y(0)}+
        \beta_5^{Y} AGE0_i^{Y(0)} + {u}_i^{Y} \\
    \frac{\partial Y_i(t)}{\partial t}  =& \gamma_0^{Y} +  \gamma_1^{Y} EL_i^{Y} + \gamma_2^{Y} APOE4_i^{Y} +
        \gamma_3^{Y} SEX_i^{Y} +\\
        &
        \gamma_4^{Y} CENTRE_i^{Y}+
        \gamma_5^{Y} AGE0_i^{Y} 
     +v_i^{Y} + \alpha^{YM}M_i(t)   + \alpha^{YL}L_i(t)  
    \end{aligned}
    \end{array}\right.
    \label{M}
\end{align}

\textbf{Study 2: effect of APOE4 on the level of verbal fluency}
 \begin{align}
&\text{For process } \mathcal{L}: \left\{
    \begin{array}{ll}
    \begin{aligned}
        L_i(0)   = &\beta_0^{L}+ \beta_1^{L} APOE4_i^{L(0)} + \beta_2^{L}AGE0_i^{L(0)} +
        \beta_3^{L} EL_i^{L(0)}  +\\
        &
        \beta_4^{L}SEX_i^{L(0)} +
        \beta_5^{L}VTI_i^{L(0)}  +
        \beta_6^{L}CENTRE_i^{L(0)} +
        {u}_i^{L} \\
    \frac{\partial L_i(t)}{\partial t} =& \gamma_0^{L} + 
    (\gamma_1^{L} APOE4_i^{L} + \gamma_2^{L} AGE0_i^{L} +
        \gamma_3^{L} EL_i^{L} +\\
        &
        \gamma_4^{L}  SEX_i^{L}+
        \gamma_5^{L} VTI_i^{L} +
        \gamma_6^{L} CENTRE_i^{L}) \times $time_{i}$ + v_i^{L}
    \end{aligned}
    \end{array}\right.\\
    &\text{For process } \mathcal{M}: \left \{
    \begin{array}{ll}
    \begin{aligned}
         M_i(0)  & = \beta_0^{M}+ \beta_1^{M} APOE_i^{M(0)} + \beta_2^{M} AGE0_i^{M(0)} +
        \beta_3^{M} EL_i^{M(0)} +\\
        &
        \beta_4^{M}  SEX_i^{M(0)}+
        \beta_5^{M} CENTRE_i^{M(0)} + {u}_i^{M} \\
    \frac{\partial M_i(t)}{\partial t} =& \gamma_0^{M} +  (\gamma_1^{M} APOE_i^{M} + \gamma_2^{M} AGE0_i^{M} +
        \gamma_3^{M} EL_i^{M} +\\
        &
        \gamma_4^{M}  SEX_i^{M}+
        \gamma_5^{M} CENTRE_i^{M}) \times $time_i$ +v_i^{M}  + \alpha^{ML}L_i(t)
    \\  
    \end{aligned}
    \end{array}\right.    \label{mo}\\
    & \text{For process } \mathcal{Y}: \left\{
    \begin{array}{ll}
    \begin{aligned}
          Y_i(0)  & = \beta_0^{Y}+ \beta_1^{Y} APOE_i^{Y(0)} + \beta_2^{Y} AGE0_i^{Y(0)} +
        \beta_3^{Y} EL_i^{Y(0)} +\\
        &
        \beta_4^{Y}  SEX_i^{Y(0)}+
        \beta_5^{Y} CENTRE_i^{Y(0)} + {u}_i^{Y} \\
    \frac{\partial Y_i(t)}{\partial t}  =& \gamma_0^{Y} +  (\gamma_1^{Y} APOE_i^{Y} + \gamma_2^{Y} AGE0_i^{Y} +
        \gamma_3^{Y} EL_i^{Y} +\\
        &
        \gamma_4^{Y} SEX_i^{Y}+
        \gamma_5^{Y} CENTRE_i^{Y}) \times $time_i$ 
     +v_i^{Y} + \alpha^{YM}M_i(t)   + \alpha^{YL}L_i(t)  
    \end{aligned}
    \end{array}\right.
    \label{M}
\end{align}

\subsection{Results of the working models}

The estimates of the two working models are reported in Tables \ref{tab:illust1} and \ref{tab:illust2} for study 1 and 2, respectively.

\begin{table}[H]
\centering
\caption{Illustration 1 : Model estimation parameters}

\label{tab:illust1}
\begin{tabular}{@{}lll@{}}
\toprule
Variable               & Coefficient & p-value \\ \midrule
$\beta_0^{L}$     & -2.7597     & 2e-16 *** \\
$\beta_1^{L}$               & -0.2649     & 0.0004 *** \\
$\beta_2^{L}$         & 0.0719      & 0.4039  \\
$\beta_3^{L}$ & 0.8075      & 2e-16 *** \\
$\beta_4^{L}$        & 0.0719      & 0.4039    \\
$\beta_5^{L}$   & 1.0284      & 2e-16 *** \\

$\beta_0^{M}$    & -0.0257     & 0.7100  \\
$\beta_1^{M}$           & 0.3723      & $<$2e-16 ***\\
$\beta_2^{M}$        & -0.0249     & 0.5090  \\
$\beta_3^{M}$  & 0.0853      & 0.0076 **  \\
$\beta_4^{M}$    & 0.1316      & 0.0001 ***    \\
$\beta_5^{M}$   & 0.1522      & 0.0005 ***  \\

$\beta_0^{Y}$    & -1.0434     & $<$2e-16 *** \\
$\beta_1^{Y} $   & 0.0495      & 0.4149 \\
$\beta_2^{Y} $    & -0.0276     & 0.7015  \\
$\beta_3^{Y} $  & -0.1525     & 0.0105 *  \\
$\beta_4^{Y}  $  & 0.5056      & $<$2e-16 ***    \\
$\beta_5^{Y}  $ & -1.8127     & $<$2e-16 ***  \\

\gamma_0^{L}    & 0.9488      & $<$2e-16 *** \\
\gamma_1^{L}     & 0.1107      & 0.0264 * \\
\gamma_2^{L}        & 0.0117      & 0.8403      \\
\gamma_3^{L}   & -0.0850     & 0.0968  \\
\gamma_4^{L}      & -0.4813     & $<$2e-16 ***   \\
\gamma_5^{L}   & 0.1848      & 0.0024 ** \\

\gamma_0^{M}  & -0.8743     & $<$2e-16 *** \\
\gamma_1^{M}        & -0.0761     & 0.0005 ***\\
\gamma_2^{M}      & -0.0616     & 0.0188 *  \\
\gamma_3^{M}   & 0.0996      & $<$2e-16 ***  \\
\gamma_4^{M}    & 0.3830      & $<$2e-16 ***  \\
\gamma_5^{M}    & -0.1865     & $<$2e-16 *** \\

\gamma_0^{Y}    & 1.6950      & $<$2e-16 ***  \\
\gamma_1^{Y}    & 0.0776      & 0.1865  \\
\gamma_2^{Y}      & 0.0183      & 0.8030 \\
\gamma_3^{Y}  & 0.0839      & 0.1538  \\
\gamma_4^{Y}     & -0.5950     & $<$2e-16 ***\\
\gamma_5^{Y}  & 0.9941      & $<$2e-16 ***   \\

Chol.1                 & 0.9733      & $<$2e-16 *** \\
Chol.7                 & 0.4418      & $<$2e-16 *** \\
Chol.16                & 0.3229      & $<$2e-16 *** \\
Chol.19                & -0.2075     & $<$2e-16 *** \\
Chol.21                & 0.6315      & $<$2e-16 *** \\

$\alpha_{YL}$.(Intercept)       & 0.2158      & $<$2e-16 *** \\
$\alpha_{YL}$.EL                & 0.0242      & 0.4699     \\
$\alpha_{YM}$.(Intercept)       & -0.7694     & $<$2e-16 *** \\
$\alpha_{YM}$.EL                & 0.1185      & 0.0911 .   \\

$\sigma_L$         & 0.9936      & $<$2e-16 *** \\
$\sigma_M$          & 0.3921      & $<$2e-16 *** \\
$\sigma_Y$         & 1.0402      & $<$2e-16 *** \\
\bottomrule
\end{tabular}
\end{table}

\begin{table}[htbp]
\centering
\caption{Illustration 2 : Model estimation parameters}
\label{tab:illust2}
\begin{tabular}{@{}lll@{}}
\toprule
Variable               & Coefficient & p-valeur \\ \midrule
$\beta_0^L$      & 9.7632      & $<$2e-16 *** \\
$\beta_1^L$            & 0.0619      & 0.1827     \\
$\beta_2^L$             & -0.1373     & $<$2e-16 *** \\
$\beta_3^L$              & 0.0832      & 0.0350 *   \\
$\beta_4^L$              & 0.1994      & 0.0001 *** \\
$\beta_5^L$              & 0.8815      & $<$2e-16 *** \\
$\beta_6^L$           & -0.2285     & $<$2e-16 *** \\

$\beta_0^M$      & 0.6917      & 0.7967     \\
$\beta_1^M$             & -0.0916     & 0.4419     \\
$\beta_2^M$             & 0.0086      & 0.8295     \\
$\beta_3^M$              & 0.0727      & 0.4672     \\
$\beta_4^M$              & -0.1691     & 0.0972 .   \\
$\beta_5^M$           & 0.3357      & 0.0041 **  \\

$\beta_0^Y$      & 5.8808      & $<$2e-16 *** \\
$\beta_1^Y$             & -0.0923     & 0.2552     \\
$\beta_2^Y$             & -0.0947     & $<$2e-16 *** \\
$\beta_3^Y$              & 0.5535      & $<$2e-16 *** \\
$\beta_4^Y$              & 0.1088      & 0.1083     \\
$\beta_5^Y$           & 0.2361      & 0.0017 **  \\

$\gamma_0^L$     & -0.8750     & $<$2e-16 *** \\
$\gamma_1^L$            & 0.0013      & 0.8944     \\
$\gamma_2^L$            & 0.0155      & $<$2e-16 *** \\
$\gamma_3^L$             & -0.0130     & 0.0997 .   \\
$\gamma_4^L$             & -0.0301     & 0.0024 **  \\
$\gamma_5^L$             & -0.0070     & 0.2166     \\
$\gamma_6^L$          & -0.0488     & $<$2e-16 *** \\
$\gamma_7^L$ x time      & -0.0569     & $<$2e-16 *** \\
$\gamma_8^L$ x time & -0.0007  & 0.4670     \\
$\gamma_9^L$x time & 0.0003   & 0.0009 *** \\
$\gamma_{10}^L$x time & 0.0013 & 0.0860 .   \\
$\gamma_{11}^L$x time & 0.0016 & 0.0970 .   \\
$\gamma_{12}^L$x time & -0.0002 & 0.7721     \\
$\gamma_{13}^L$x time & 0.0048 & $<$2e-16 *** \\

$\gamma_0^M$     & 0.2550      & 0.4145     \\
$\gamma_1^M$            & 0.0145      & 0.6279     \\
$\gamma_2^M$            & -0.0013     & 0.7815     \\
$\gamma_3^M$& -0.0130     & 0.5909     \\
$\gamma_4^M$             & 0.0207      & 0.4175     \\
$\gamma_5^M$          & -0.1037     & 0.0002 *** \\
$\gamma_6^M$    & 0.0689      & 0.0227 *   \\
$\gamma_7^M$x time & 0.0005    & 0.8839     \\
$\gamma_8^M$x time & -0.0007   & 0.0668 .   \\
$\gamma_9^M$x time & -0.0004    & 0.8722     \\
$\gamma_{10}^M$x time & -0.0021 & 0.4511     \\
$\gamma_{11}^M$x time & -0.0022 & 0.4670     \\

\bottomrule
\end{tabular}
\end{table}

\begin{table}[H]
\centering

\label{tab:my-table}
\begin{tabular}{@{}lll@{}}
\toprule
Variable               & Coefficient & p-valeur \\ \midrule
$\gamma_0^Y$     & -0.5890     & $<$2e-16 *** \\
$\gamma_1^Y$            & 0.0021      & 0.8731     \\
$\gamma_2^Y$            & 0.0079      & $<$2e-16 *** \\
$\gamma_3^Y$             & -0.0163     & 0.0963 .   \\
$\gamma_4^Y$            & 0.0049      & 0.6327     \\
$\gamma_5^Y$          & 0.0526      & $<$2e-16 *** \\
$\gamma_6^Y$     & 0.0125      & 0.0927 .   \\
$\gamma_7^Y$x time & -0.0005  & 0.6594     \\
$\gamma_8^Y$x time& -0.0002  & 0.0139 *   \\
$\gamma_9^Y$x time & 0.0005    & 0.5104     \\
$\gamma_{10}^Y$x time & 0.0003  & 0.6524     \\
$\gamma_{11}^Y$x time& -0.0022 & 0.0060 **  \\
Chol.1                 & 0.4333      & $<$2e-16 *** \\
Chol.2                 & -0.1893     & $<$2e-16 *** \\
Chol.3                 & 0.0800      & 0.0002 *** \\
Chol.4                 & -0.0036     & 0.1304     \\
Chol.7                 & 1.0059      & $<$2e-16 *** \\
Chol.8                 & -0.0268     & 0.2392     \\
Chol.10                & 0.0763      & 0.0021 **  \\
Chol.12                & 0.7667      & $<$2e-16 *** \\
Chol.15                & -0.0220     & $<$2e-16 *** \\
Chol.16                & -0.0270     & $<$2e-16 *** \\
Chol.19                & -0.0499     & 0.4019     \\
Chol.21                & 0.0373      & $<$2e-16 *** \\
$\alpha_{ML}.(Intercept)$       & -0.0002     & 0.9758     \\
$\alpha_{ML}.APOE $             & 0.0019      & 0.8678     \\
$\alpha_{YM}.(Intercept)$       & -0.0031     & 0.0029 **  \\
$\alpha_{YM}.APOE$              & -0.0011     & 0.5614     \\

$\sigma_L$               & -0.1774     & $<$2e-16 *** \\
$\sigma_M$              & 0.4719      & $<$2e-16 *** \\
$\sigma_Y$              & 0.5059      & $<$2e-16 *** \\ \bottomrule
\end{tabular}
\end{table}